# Doubts on the efficacy of outliers correction methods

**Marjorie Fonnesu and Nicola Kuczewski**

Lyon Neuroscience Research Center, University Lyon1, CNRS, UMR 5292; INSERM, U1028.

Address correspondence to: Nicola.kuczewski@ubiv-lyon1.fr

## Abstract

While the utilisation of different methods of outliers correction has been shown to counteract the inferential error produced by the presence of contaminating data not belonging to the studied population; the effects produced by their utilisation when samples do not contain contaminating outliers are less clear. Here a simulation approach shows that the most popular methods of outliers correction (2 Sigma, 3 Sigma, MAD, IQR, Grubbs and winsorizing) worsen the inferential evaluation of the studied population in this condition, in particular producing an inflation of Type I error and increasing the error committed in estimating the population mean and STD. We show that those methods that have the highest efficacy in counteract the inflation of Type I and Type II errors in the presence of contaminating outliers also produce the stronger increase of false positive results in their absence, suggesting that the systematic utilisation of methods for outliers correction risk to produce more harmful than beneficial effect on statistical inference. We finally propose that the safest way to deal with the presence of outliers for statistical comparisons is the utilisation of non-parametric tests.

## Introduction

Outliers have been defined as observations that "deviate so much from other observations as to arouse suspicion that it was generated by a different mechanism"(Hawkins, 1980) or observations that "appears to be inconsistent with the remainder of that set of data"(Barnett & Lewis, 1994)". The first point highlighted by these definitions is that outliers are generally considered as contaminating data that do not belong to the population of the parameter that is under investigation. As a consequence, the presence of outliers in the analysed sample is seen as an external element worsening the inferential evaluation of the studied population. The second point highlighted by the aforementioned definitions is that the outliers can be identified only relatively to the available data sample; outlier detection is therefore submitted to the inferential errors linked to random sampling. Several methods of outliers detection have been developed among which the most popular are the following: ***a)*** the mean plus or minus two-three standard deviation (Sigma 2 and Sigma 3) methods, ***b)*** the interquartile range (IQR) method, ***c)*** the absolute deviation around the median (MAD) method, ***d)*** the Grubbs method and ***e)*** the trimming method. (Barnett & Lewis, 1994; Seo, 2006; Leys *et al.*, 2019) for more details on the different methods. Once detected outliers are in general removed for the sample or transformed to be reduced in the limit of the acceptable data, a procedure known as winsorizing. The efficacy of different outliers correction methods has been evaluated and compared mainly by approaches using simulation of data sets (Zimmerman, 1994, 1995; Seo, 2006; Liao *et al.*, 2016, 2017) or by randomly extracting real data from a known population arbitrary divided in 'Normal' and 'Outliers' subjects (Osborne & Overbay, 2004). These approaches have demonstrated that the presence of contaminating outliers in the analysed sample worsen the inferential evaluation of population, producing an inflation of Type I and Type II error as well as an increase of the error committed in estimating the population effect. They have also highlighted the benefit of the utilisation of the different correction methods in contaminated sample, that in several cases counteract the negative effect produced by the presence of the outliers. While these approaches have focalised their attention to the effect produced by outliers correction in the case where outliers are present in the analysed samples, they do not evaluate the impact produced by the application of these methods on samples that do not contain external contaminating data. Indeed, it is conceivable that some of the data belonging to the studied population would be considered as outliers when screened by the different methods; this as a consequence of the rule of detection (ex. Sigma 2 methods would consider outliers ~5% of the data coming from normally distributed population) as well as a consequence of the inferential error discussed above. From now on such outliers produced by inherent variability will be designated as random sampling outliers (RSO). Given that the presence of RSO in the data set cannot be discriminate from contaminating outliers, the utilisation of outlier correction methods could in several circumstance target mainly or exclusively the former. By using the mean plus or minus sigma method on simulated data coming from normal distribution, Bakker and Wicherts showed that in the absence of contaminating outliers, the removal of RSO produce an increase of the occurrence of Type I when the p-value was evaluated with a t-test statistic (Bakker & Wicherts, 2014). While this report highlight the possible risk of outlier removal the conclusions are limited to a specific experimental condition. The main objective of the present work was therefore to extend the investigation on the impact produced by the correction of RSO on inferential statistic. To this goal we use an approach based on simulated data to address the following questions: 1-What is the proportion of RSO detectable as a function of correction method, samples size and type of data distribution? 2-Does the presence of RSO in the analysed samples affect the inferential evaluation of the studied population? 3-What is the impact of the correction of RSO on inferential statistic? In particular, on the evaluation of population parameters (mean and std), on the occurrence of Type I and Type II error and on the estimation of the real effect when samples belong to different populations. 4- Does an eventual impact of RSO correction on Type I and Type II error depends on the type of statistical test used to compare the samples? 5- Do the different methods of outliers correction have similar impact on inferential statistic depending whether samples present RSO or contaminating outliers?

An important difference compared to Bakker and Wicherts 2014, is the fact that in the present work the average impact produced by the correction of RSO is evaluated on the selected samples population containing RSO and not on the whole samples population. We believe that this approach is more informative for a researcher that, facing to an experimental sample for which some of the data can be labelled as outliers, has to evaluate the risk-benefit ratio of their correction.

## Methods

Data simulations were performed with Python 3.7. np.random.normal and np.random.lognormal function were used to generate 100000 random samples, or pairs of samples, for each condition. This number was empirically determined by the procedure depicted in supplementary figure 1. A similar approach was used to determine the number of permutations (600) performed for the permutation test (not shown). These simulations were performed from samples coming from normal and log-normal distributions having $\mu=0$ and $\sigma=1$. It should be pointed out that for log-normal distribution $\mu$ and $\sigma$ are not the values for the distribution itself but that of the underlying normal distribution it is derived from (random.lognormal function from numpy), what give a real $\mu \simeq 1.649$ and real $\sigma \simeq 2.16$. Outlier correction methods: Grubbs was imported from Smirnov_grubbs package; Winsorizing was performed with mstats.winsorize with 5% winsorisation; Median absolute deviation used for MAD method was performed with mad_std function (astropy.stats), values x where considered to be outliers when $|(x-Median(sample))|/MAD>2.24$. In some of the simulations the more restrictive threshold of 3 for MAD detection was used (Ley et al 2013). In Winsorizing sigma 2 method data above 2 STD from sample mean were reduced to 2 STD value. All the scripts are available at the Open Science Framework (https://osf.io/gzra8/).

## Results

### Part 1: Random Sampling Outlier (RSO)

Here RSO are defined as the data present in a sample that, despite belonging to the studied population, are considered to be outliers by the different detection methods. In these simulations therefore all the data are sampled from the target distributions, i.e. no external contaminating data are injected in the samples. It should be stressed the fact that some of the data can be labelled as outliers with respect to a specific sample without necessary be extreme values of the studied population. In the same way extreme values of the studied population could be considered as normal value with respect of a specific sample (some examples of these cases are presented in supplementary figure 4).

#### Simulation 1: Impact of RSO correction on the estimation of population parameters

Simulation 1 was performed by the procedure descripted in Figure 1 (the script is available at https://osf.io/753yc/). This allows to extract, for the different detection methods, 10000 samples of sizes **n** having at least one RSO in them. The error in the estimation of population mean and standard deviation was expressed in standard deviations of the population distribution.

The objectives of this simulation were:

1- Determine the proportion of RSO detected by the different methods for different sample sizes
2- Determine whether and how the correction of RSO from the samples modify the estimation of population average and standard deviation

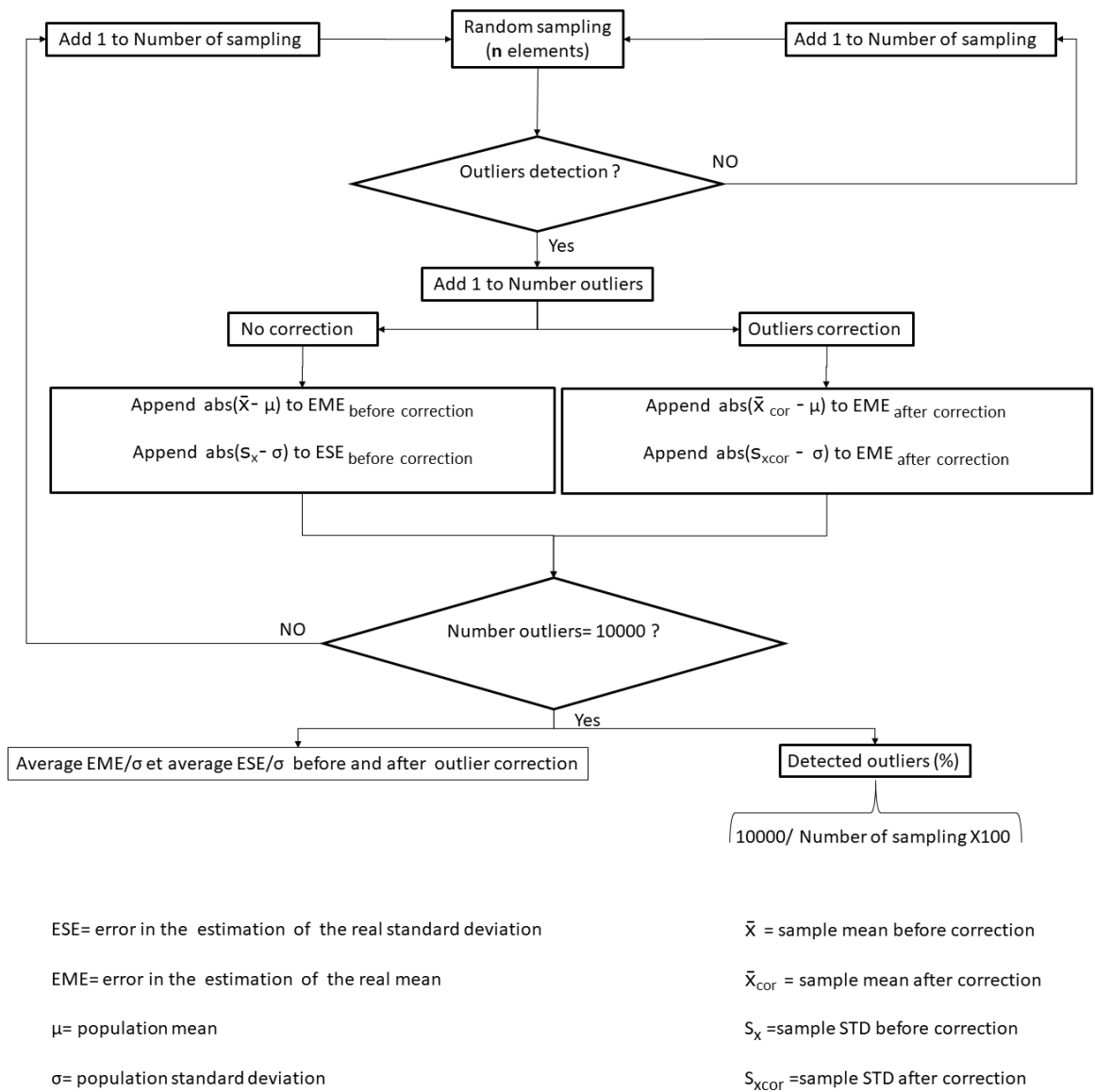

**Fig 1: Procedure used to simulate the effect of the correction of "Random Sampling Outliers" on the estimation of real population mean and standard deviation.** This procedure was separately applied for the different methods of outliers correction, sample size and population distribution (Normal or Log-normal). The error is expressed in population STD. "Number of sampling "and "Number of outliers "were set to zero at the beginning of the simulation (script is available at https://osf.io/753yc/).

*Correction of RSO increase the error in the estimation of population mean and std (samples from normal distribution)*

The first objective was to determine the probability to observe at least one RSO in a random sample from normally distributed population (the RSO probability) as a function of sample size and outliers detection method (Figure 2 red lines). With the exception of "Grubbs" methods the RSO probability increase with the sample size eventually becoming 100% for large sample. As a consequence the difference in RSO probability between different methods is principally observed for the smallest sample (n<100-500), where it is higher when IQR and MAD method are used. "Grubbs" method is particularly inefficient in detecting RSO, especially when sample size increase. It is interesting to note that RSO detectable by "Sigma 3" method are never present when sample size is ≤11 (Fig 2C), such results is in agreement with theoretical prediction (Shiffler, 1988). The second objective was to determine whether RSO correction impact the estimation of real population mean ($\mu$) and STD($\sigma$). As show in figure 2A and 3A, when samples are not selected based on the presence of outliers the estimation error for the two parameters decrease with the sample size. Similar decrease is observed when samples are selected for the presence of RSO (Fig 2 B-H and 3B-H blue lines). Outliers correction never reduce the average error in estimating **μ** (Fig 2) and **σ** (Fig 3). On the contrary most of the correction methods worsen the estimation of this two parameters, especially for the smallest sample size (n<50). The only exception is the "Accommodation sigma 2" method for which outliers correction do not modify the error in estimating **μ** and produce only a slight increase of the error in **σ** estimation. Supplementary figure 2 allow to compare the impact of "Sigma 2" and "Accommodation sigma 2" methods on the **μ** and **σ** estimation error distributions when sample size =6. It also shows an example of sample correction that brings to an increase of **μ** and **σ** evaluation error when "Sigma 2" correction method is used but no relevant change on the evaluation of these two parameters when "Accommodation sigma 2" method is used. Readers interested to run these simulations with other parameters are invited to use the following scripts: https://osf.io/dfeas/ for the error distributions, https://osf.io/t9mvc/ for sample correction example.

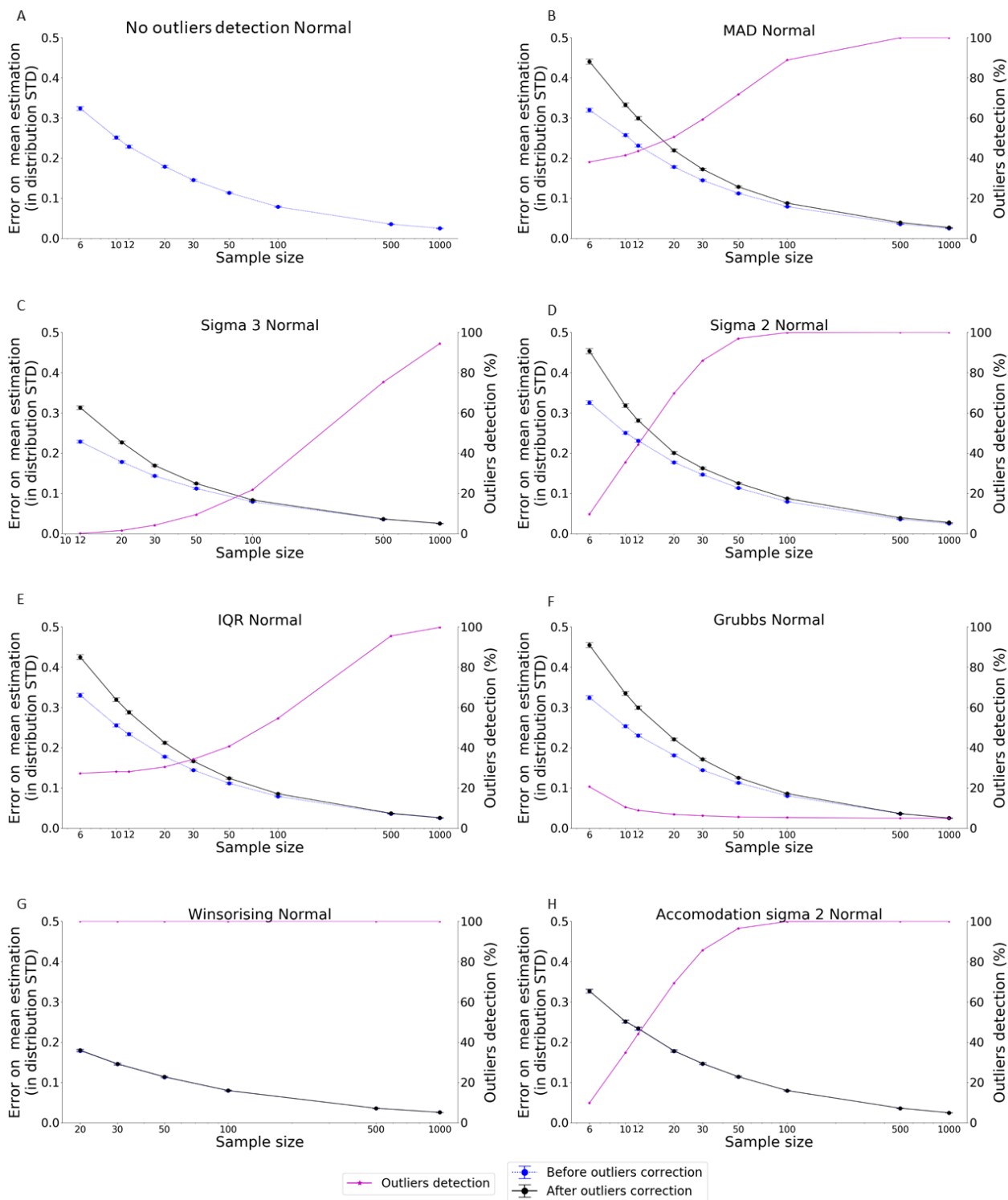

**Fig 2: Outlier correction (RSO) increase the average error on population mean estimation (data from normal distribution).** In A samples are not selected based on the presence of outliers. Note that in C and G sample size start from 12 and 20 respectively. Vertical bars represent 95% CI. All data can be found at https://osf.io/pjcak/

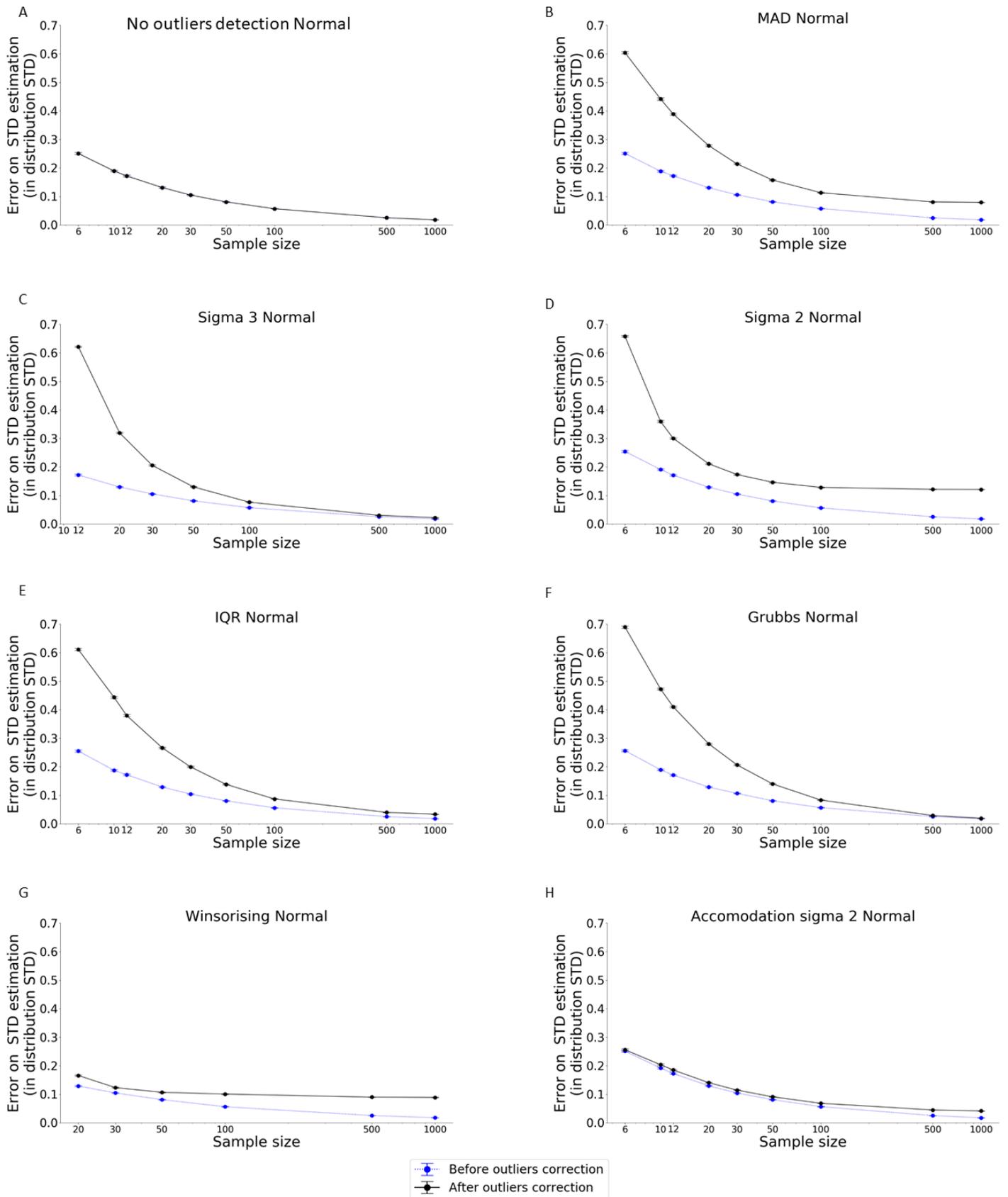

**Fig 3: Outlier correction (RSO) increase the error on population STD estimation (data from normal distribution).** In A samples are not selected based on the presence of outliers. Note that in C and G sample size start from 12 and 20 respectively. Vertical bars represent 95% CI. All data can be found at https://osf.io/pjcak/

*Correction of RSO increase the error in the estimation of population mean and std (samples from log-normal distribution)*

Compared to normal distribution the probability that a sample data would be considered as an outlier is greatly increased when samples came from log-normal distribution. The RSO probability increase with the sample size for all tested detection methods and, when n=100, become 100% whatever methods is used (Fig 4, red line). Even in this case no data further than 3 STD from sample mean are observed when sample size is ≤11 (Fig 4 C). The error in estimating **μ** and **σ** from sample mean and sample STD is larger compared to sample coming from normal distribution but still decrease with sample size (Fig 4A and Fig 5A). Again outliers correction worsen the estimation of **μ** and **σ** but in this case the worsening increase with the sample size. It should be noted that when "Accommodation sigma 2" method is used a small decreased of **μ** estimating error is observed for sample size between 6 and 30 (Fig 5H). Supplementary figure 3 allow to compare the impact of "Sigma 2" and "Accommodation sigma 2" methods on the **μ** and **σ** estimation error distributions when sample size =100. It also shows un example of sample correction that brings to an increase of **μ** and **σ** evaluation error when "Sigma 2" correction method is used but only a relevant change on the evaluation of **σ** when "Accommodation sigma 2" method is used.

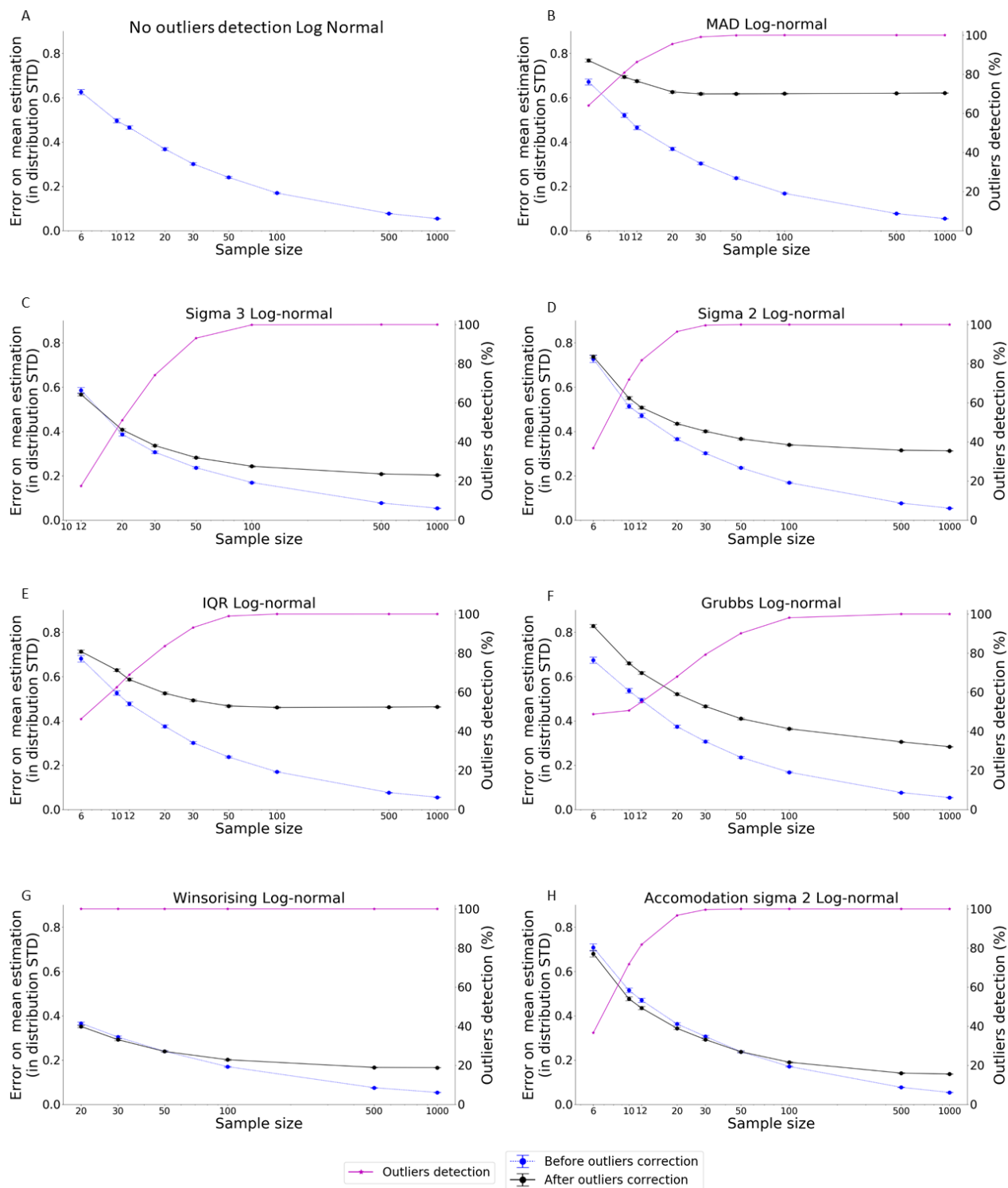

**Fig 4: Outlier correction (RSO) increase the average error on population mean estimation (data from log normal distribution).** In A samples are not selected based on the presence of outliers. Note that in C and G sample size start from 12 and 20 respectively. Vertical bars represent 95% CI. All data can be found at https://osf.io/gq63d/

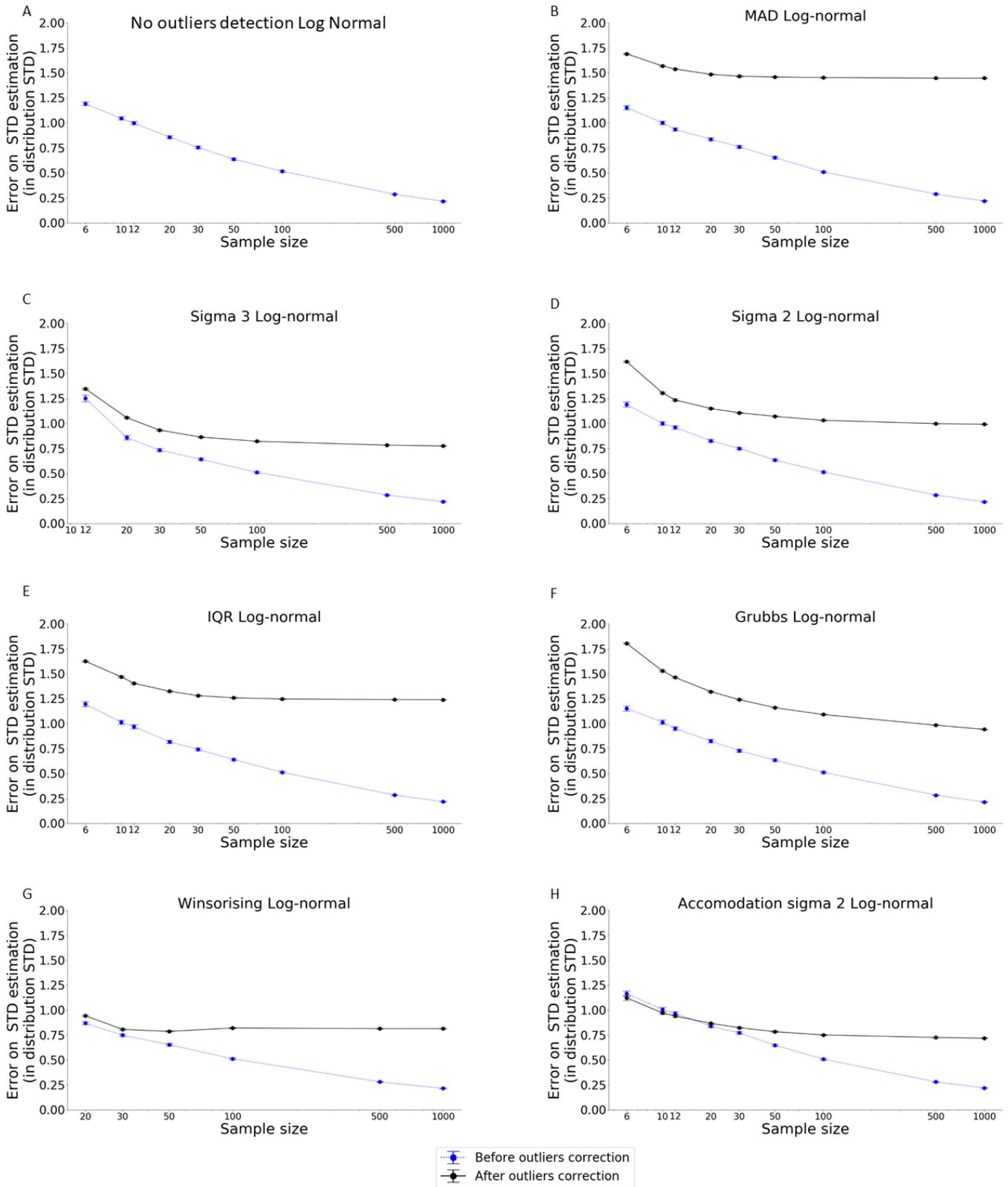

**Fig 5: Outlier correction (RSO) increase the average error on population STD estimation (data from log normal distribution).** In A samples are not selected based on the presence of outliers. Note that in C and G sample size start from 12 and 20 respectively. Vertical bars represent 95% CI. All data can be found at https://osf.io/gq63d/

*Conclusions from simulation 1*

Simulation 1 shows that:

1- When using current detection methods some of the data belonging to studied population are considered to be outliers (i.e. the tested detection methods produce RSO)
2- The probability of RSO detection increase with the sample size, depend on the detection method and is larger when samples come from log normal than normal distribution.
3- The correction of RSO (removal or winsorisation) worsens the estimation of population mean and standard deviation that can be drawn from the samples.

## Simulation 2. Impact of RSO correction on type I error

Simulation 2 was performed by the procedure descripted in Figure 6 (the script is available at https://osf.io/jbuvc/). This allow to extract, for the different detection methods, 10000 pairs of samples of sizes **n**, belonging to the same population and having at least one RSO in them. The occurrence of type I error was calculated by dividing the number of comparison, within pairs, giving a false positive (FP) result (i.e. when p-value<0.05) by the number of pair (10000). Statistical comparisons were made using Student Ttest, Mann-Whitney test and permutation test.

The objectives of this simulation were:

1- Determine whether the presence of RSO affect the occurrence of Type I error rate.
2- Determine whether and how the correction of RSO from the samples modify the occurrence of type I error.
3- Determine whether the effect of outlier correction depend on the statistical test used for samples comparison.

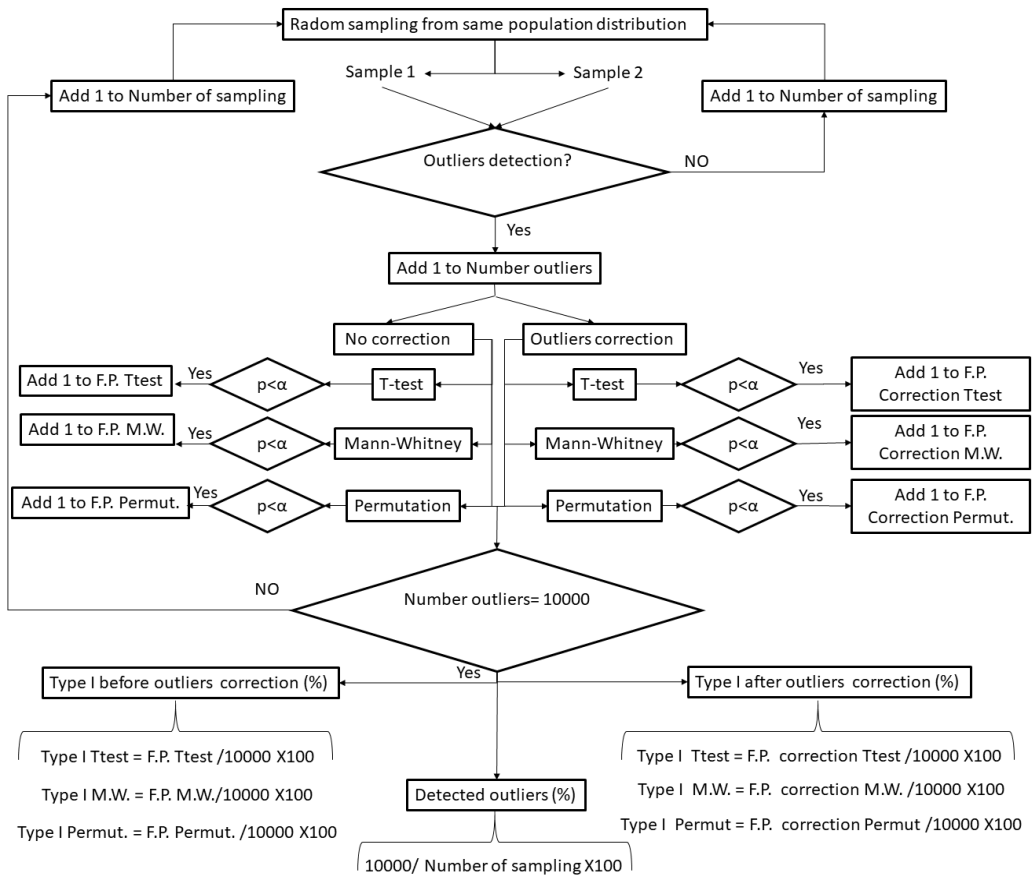

**Fig 6: Procedure used to simulate the effect of outlier correction on the occurrence of Type I error for "Random Sampling Outliers".** Outliers where separately detected for sample 1 and sample 2. Detection means at least 1 outlier in either sample 1 or sample 2. This procedure was separately applied for the different methods of outliers detection and different type of population distribution (Normal, Log-normal and Exponential). F.P.= false positives. Number of outliers, number of sampling and F.P. were set equal to 0 at the beginning of the simulation. M.W. = Mann Witney test ; Permut.= Permutation test .(Script is available at https://osf.io/jbuvc/).

*Correction of RSO increase the occurrence of Type I error (samples from normal distribution)*

As shown in figure 7 (uncoloured lines), in samples selected for the presence of RSO the Type I error is, in general, maintained at the **σ** risk (5%). The only exception is observed when RSO are detectable by "Sigma 3" or "Grubbs" methods. In these cases, an inflation of Type I error is observed, but only for sample size between 6 and 100 and when Mann Whitney test is used. Correction of RSO produce an inflation of Type I error for all correction methods except for "Accommodation sigma 2" (coloured lines). Type I error inflation after correction is higher for small sample size and is abolished for n≥ 500 when "Sigma 3" and "Grubs" methods are used. Moreover, inflation produced by outlier correction is slightly more important when Permutation and Ttest are used compared to Mann-Whitney test. Supplementary figure 4 show some examples where outliers correction generates false positive results (readers interested to run this simulation with other parameters are invited to use the following scripts: https://osf.io/zhyk3/).

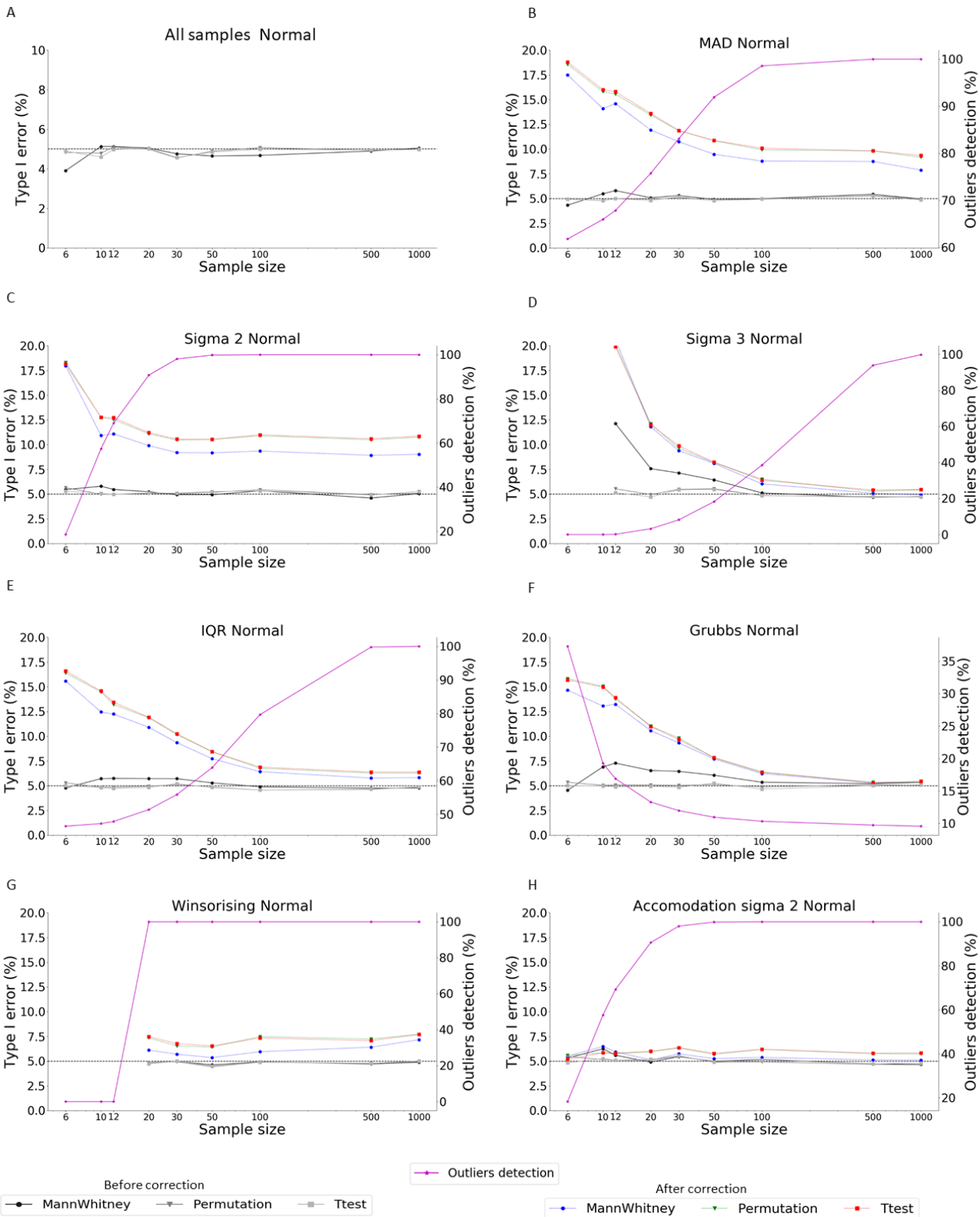

**Fig 7: Impact of correction of "Random sampling outlier" on Type I error (data from normal distribution).** In A samples are not selected based on the presence of outliers. All data can be found at: https://osf.io/cq4w3/

*Correction of RSO increase the occurrence of Type I error (Samples from log-normal population)*

As shown in figure 8 (uncoloured lines), the presence of RSO do not impact the occurrence of the Type I error, that is maintained at the σ risk (5%). However, outlier correction produces a strong inflation of Type I error, especially when Permutation test and Ttest are used (green and red lines). In these cases, the inflation of Type I error produced by the correction increase with the sample size. A notable exception is the utilization of "Accommodation sigma 2" that, used in association to Mann-Whitney test, keep the Type I error at the σ risk.

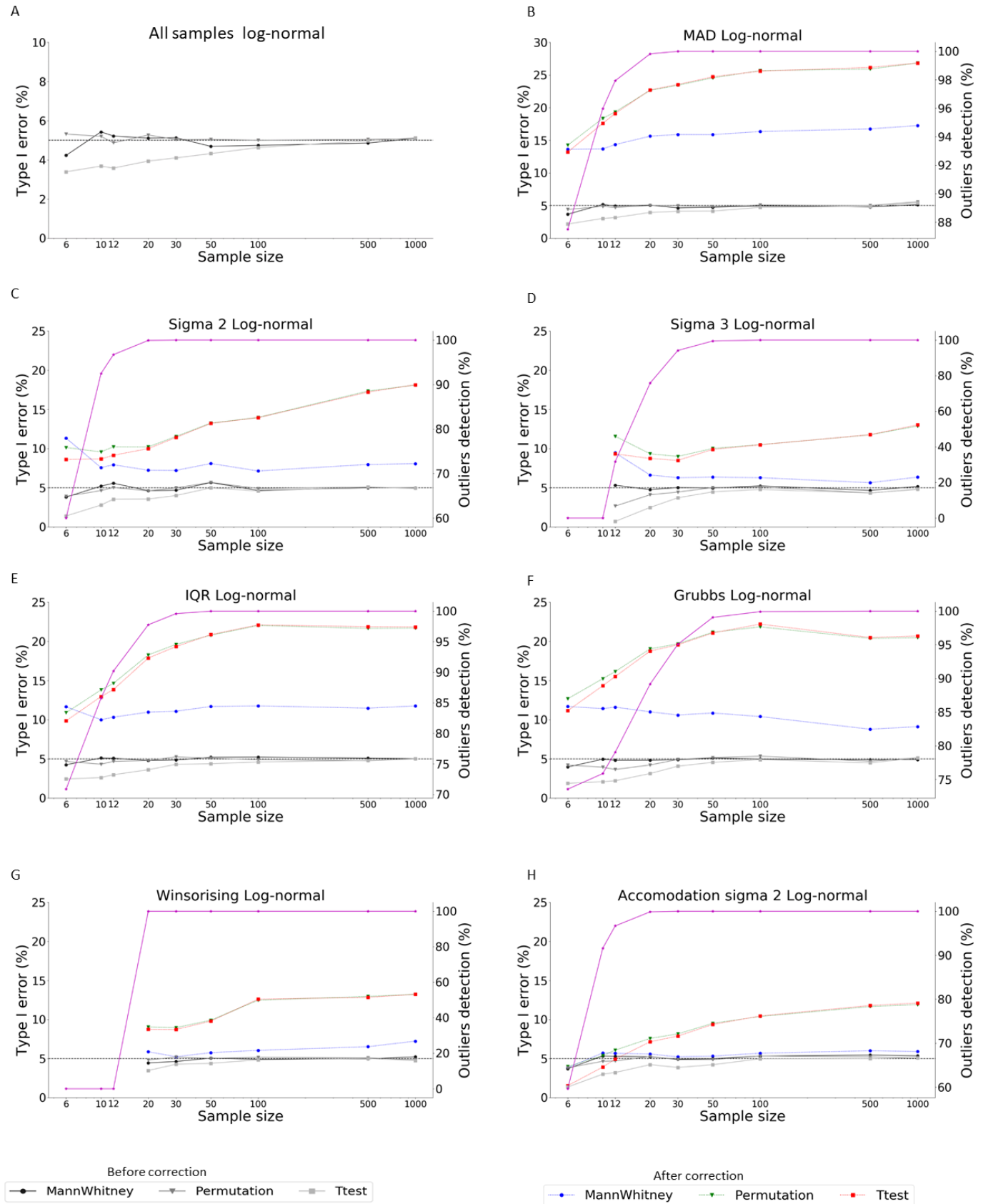

**Fig 8: Impact of correction of "Random sampling outlier" on Type I error (data from log- normal distribution).** In A samples are not selected based on the presence of outliers. All data can be found at: https://osf.io/hezmw/

We also tested winsorizing correction at 10 and 20 %. In these conditions a strong increase of type I error was produced (Supplementary figure 5).

*Conclusions from simulation 2*

The results from simulation 2 show that:

1. The presence of RSO can produce a small inflation of Type I error only in the case in which RSO are detectable with "Sigma 3 "and "Grubbs" methods, samples belong to normal distribution, have a size between 6 and 50 and comparison is made Mann-Witney test (Fig 9 D and F, black line).
2. The correction of RSO (removal or winsorisation) produces an inflation of Type I error for all tested method except when the "Accommodation sigma 2" method is used in association with Mann Whitney test.
3. The inflation of type I error produced by RSO correction is more important when samples are collected from log-normal distribution.
4. The inflation of type I error produced by RSO correction is more important when statistical comparison of the sample is made with Permutation test and Ttest compared to Mann Whitney test.

## Simulation 3. Impact of RSO correction on type II error

Simulation 3 was performed by the procedure descripted in Figure 9 (the script is available at https://osf.io/jkg75/ ). This allows to extract, for the different detection methods, 10000 pairs of samples of sizes **n**, belonging to the separate populations and having at least one RSO in them. The occurrence of type II error was calculated by dividing the number of comparison between pairs giving a false negative (FN) result (i.e. when p-value>0.05) by the number of pair (10000). Statistical comparisons were made using Student ttest, Mann-Whitney test and permutation test. In these simulations a statistical power of either 50% or 95% was obtained by the modification of the distributions mean ($\mu_2$) from which sample 2 was extracted as a function of the required sample size and distribution types. When samples came from normal distribution the value of $\mu_2$ was chosen in order to observe the target power when t-test is used. When samples came from log-normal distribution the value of $\mu_2$ was chosen in order to observe the target power when Mann-Whitney test is used (Table 1). The mean of the distribution from which sample 1 was extracted was $\mu=0$. Distributions standard deviation $\sigma=1$.

The objectives of this simulation were:

1. Determine whether the presence of RSO affect the occurrence of Type II error rate.
2. Determine whether and how the correction of RSO from the samples modify the occurrence of type II error.
3. Determine whether the effect of outlier correction on type II error depend on the statistical test used for samples comparison and statistical power.

| Sample size | µ2 for normal distribution | | µ2 for log-normal distribution | | Real µ2 of the log-normal distribution | |
|---|---|---|---|---|---|---|
| | Power 50 % | Power 95 % | Power 50 % | Power 95 % | Power 50 % | Power 95 % |
| 6 | 1.252 | 2.3 | 1.352 | 2.49 | 6.37 | 19.89 |
| 10 | 0.926 | 1.7 | 0.94 | 1.76 | 4.22 | 9.58 |
| 12 | 0.837 | 1.54 | 0.85 | 1.575 | 3.86 | 7.96 |
| 20 | 0.636 | 1.17 | 0.655 | 1.195 | 3.17 | 5.45 |
| 30 | 0.515 | 0.95 | 0.528 | 0.97 | 2.79 | 4.35 |
| 50 | 0.396 | 0.73 | 0.405 | 0.74 | 2.47 | 3.45 |
| 100 | 0.279 | 0.51 | 0.284 | 0.525 | 2.19 | 2.78 |
| 500 | 0.124 | 0.23 | 0.128 | 0.235 | 1.87 | 2.08 |
| 1000 | 0.0877 | 0.16 | 0.09 | 0.164 | 1.80 | 1.94 |

**Table 1: µ2 values of the simulated distributions allowing to have the required statistical power for a given sample size.** µ1 and Sigma were respectively set equal to 0 and 1. For normal distribution, µ2 were calculated using GPower3 (the target power is reached when t-test is used). For log-normal distributions µ2 were empirically determined in order to attend the target power when Mann-Whitney test is used.

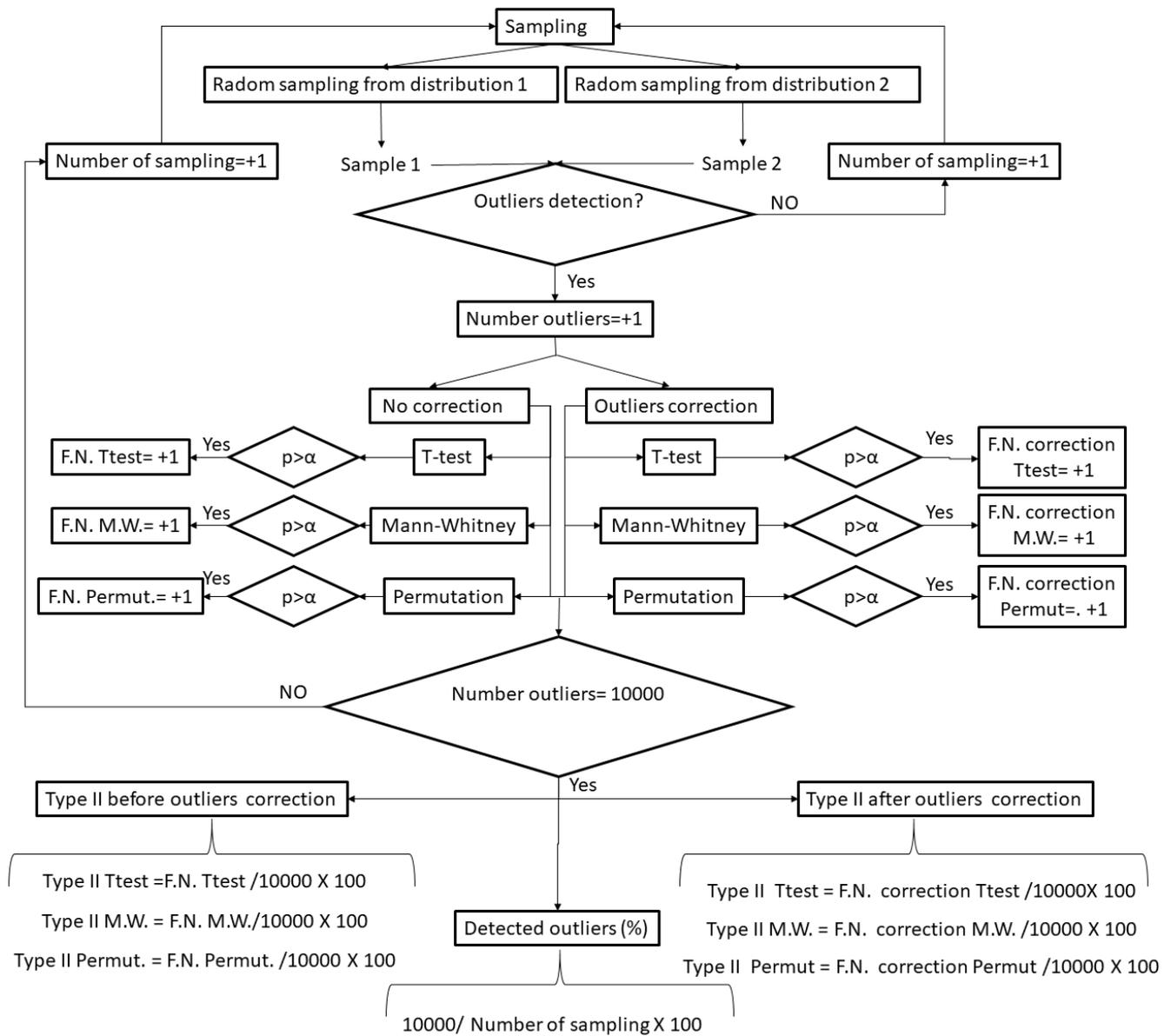

**Fig 9: Procedure used to simulate the effect of outlier correction on the occurrence of Type II error for "Random Sampling Outliers".** Outliers where separately detected for sample 1 and sample 2. Detection means at least 1 outlier in either sample 1 or sample 2. This procedure was separately applied for the different methods of outliers detection and different types of population distribution (Normal, Log-normal). F.N.= false negatives. Number of outliers, number of sampling and F.N. were set equal to 0 at the beginning of the simulation. M.W. = Mann Witney test ; Permut.= Permutation test. Script is available at https://osf.io/jkg75/.

***Correction of RSO mainly decrease the occurrence of Type II error when samples come from normal population.***

As shown in figure 10 and 11 (uncoloured lines), in samples selected for the presence of RSO the Type II error is maintained at the targets β risk (50% and 95%) when Ttest and Permutation test are used. On the other hand, RSO detectable by "Sigma 3" method reduce occurrence of false negatives when samples obtained in low powered conditions are compared with the Mann-Whitney test (Fig 10D). Outlier correction has a different effects depending on the statistical power. In low powered conditions outlier correction reduce the probability to observe false negative results, in particular when Ttest and Permutation test are used (Fig 10). In higher powered conditions outliers correction has more heterogeneous effect, producing either a reduction or an increase of type II error that depend on method and the test used (Fig 11). These modifications are extremely mild (1-2%) except when RSO are detectable by Grubss test and sample size =6, in this case a consistent increase of false positives is observed after correction (Fig 11 F).

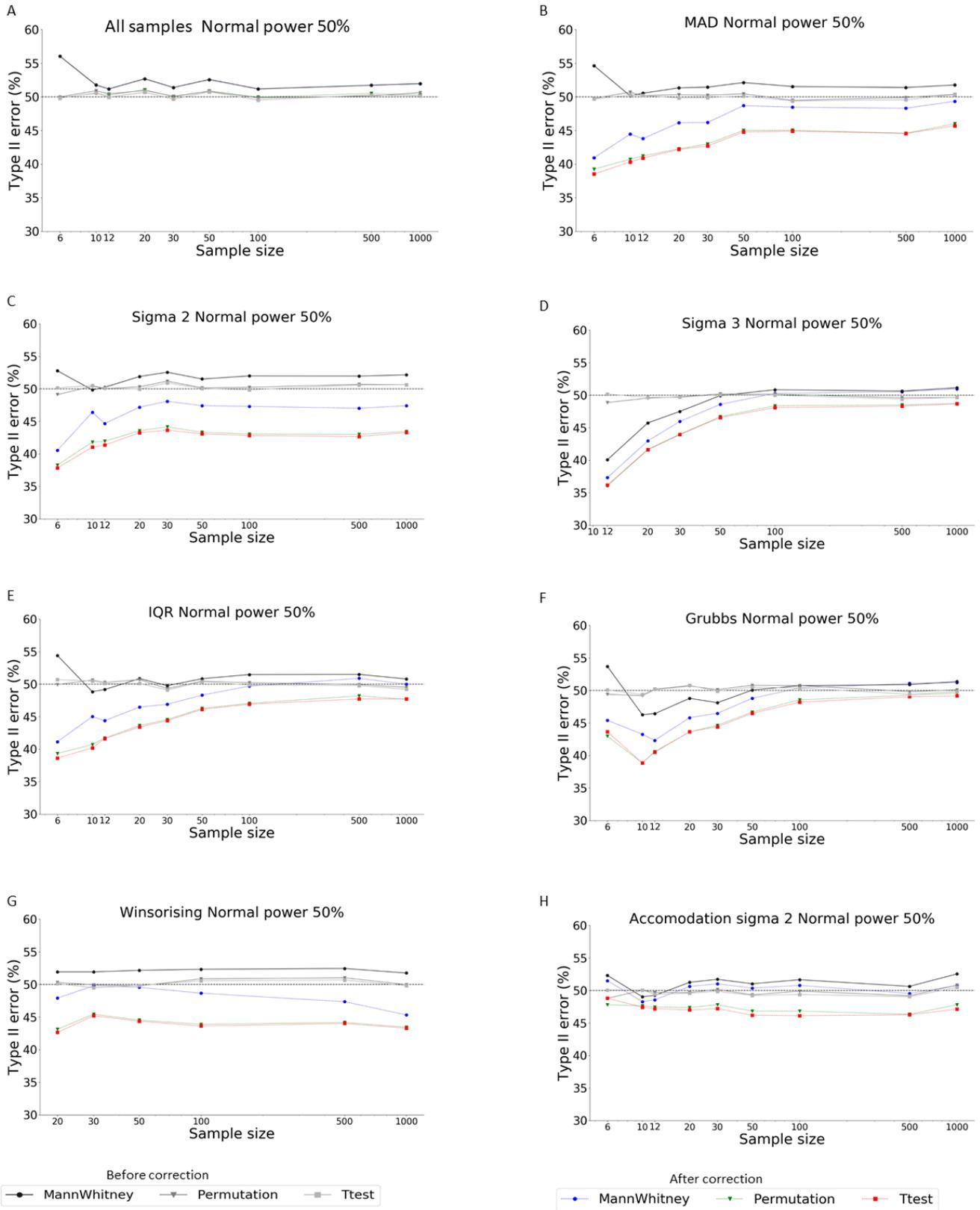

**Fig 10: Impact of correction of "Random sampling outlier" on Type II error when statistical power = 50% (data from normal distributions).** In A samples are not selected based on the presence of outliers. All data can be found at https://osf.io/p4t9q/

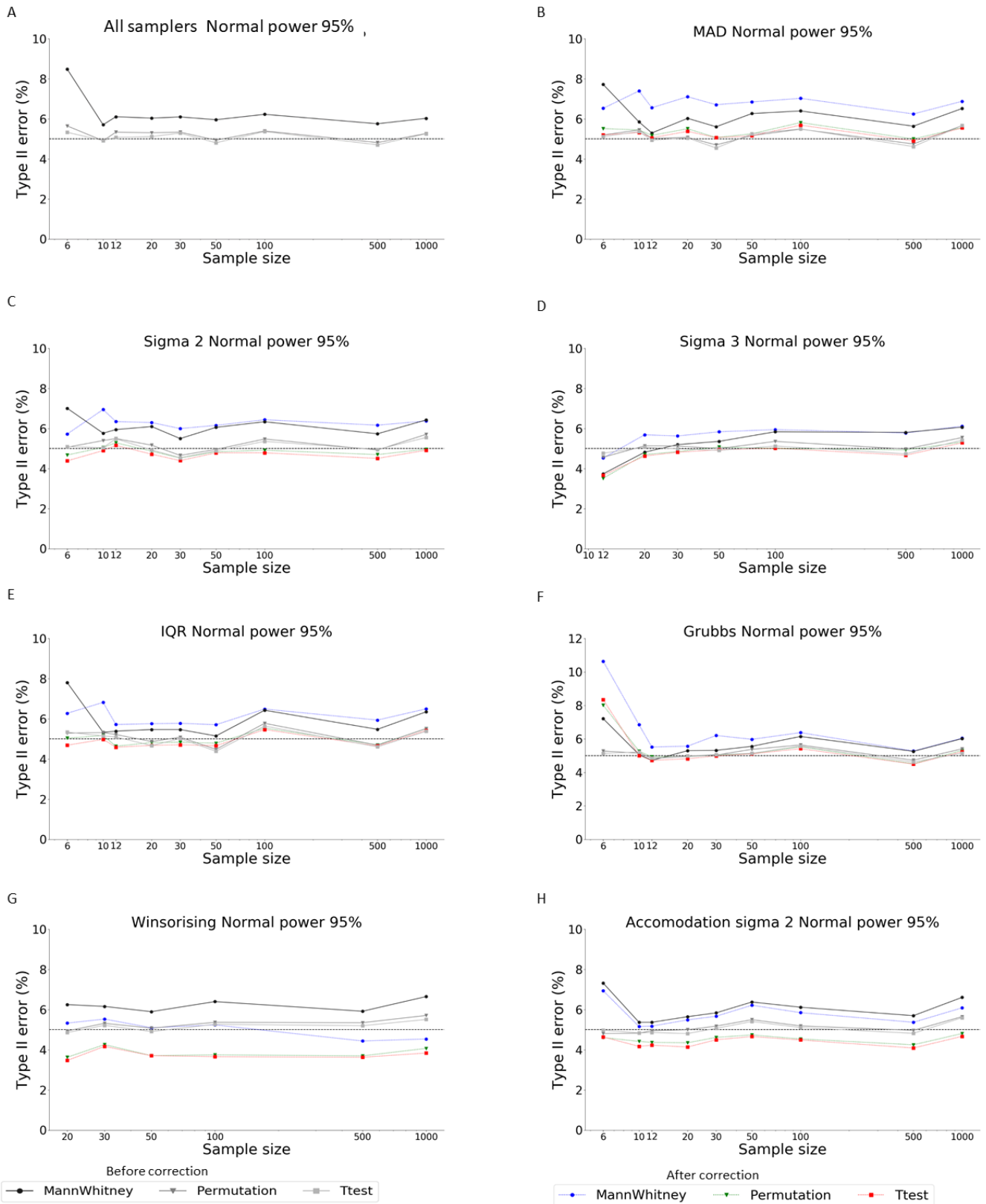

**Fig 11: Impact of correction of "Random sampling outlier" on Type II error when statistical power = 95% (data from normal distributions).** In A samples are not selected based on the presence of outliers. All data can be found at https://osf.io/p4t9q/

*Correction of RSO mainly decrease the occurrence of Type II error when samples come from log-normal population.*

Figure 12A and 13A show the inadequacy of Ttest and Permutation test to detect statistically significant differences between samples arising from populations having a log-normal distribution. A strong increase of Type II error is observed in this condition that is slightly lower when permutation test is associated to smaller sample size. In samples selected for the presence of RSO there are not relevant modification of Type II error both when using parametric and non-parametric tests, in the latter case false negative rate remain at the targets β risk (50% and 95%) (uncoloured line of Fig 12 B-H, Fig 13 B-H). Outlier correction has a beneficial impact on the occurrence of false negative results. First it strongly reduces the inflation of type II error produced by the utilisation of Permutation test and Ttets. Second, when associated with low powered comparisons, they reduce the occurrence of type II error below the target β risk. A notable exceptions are: a) the "Grubbs" method in high powered comparisons and sample size=6 for which an increase of Type II error is observed for permutation and Mann Whitney tests (Fig 13F); b) the "Accommodation sigma 2" methods associated with Mann Whitney test for which no modification of Type II error are observed.

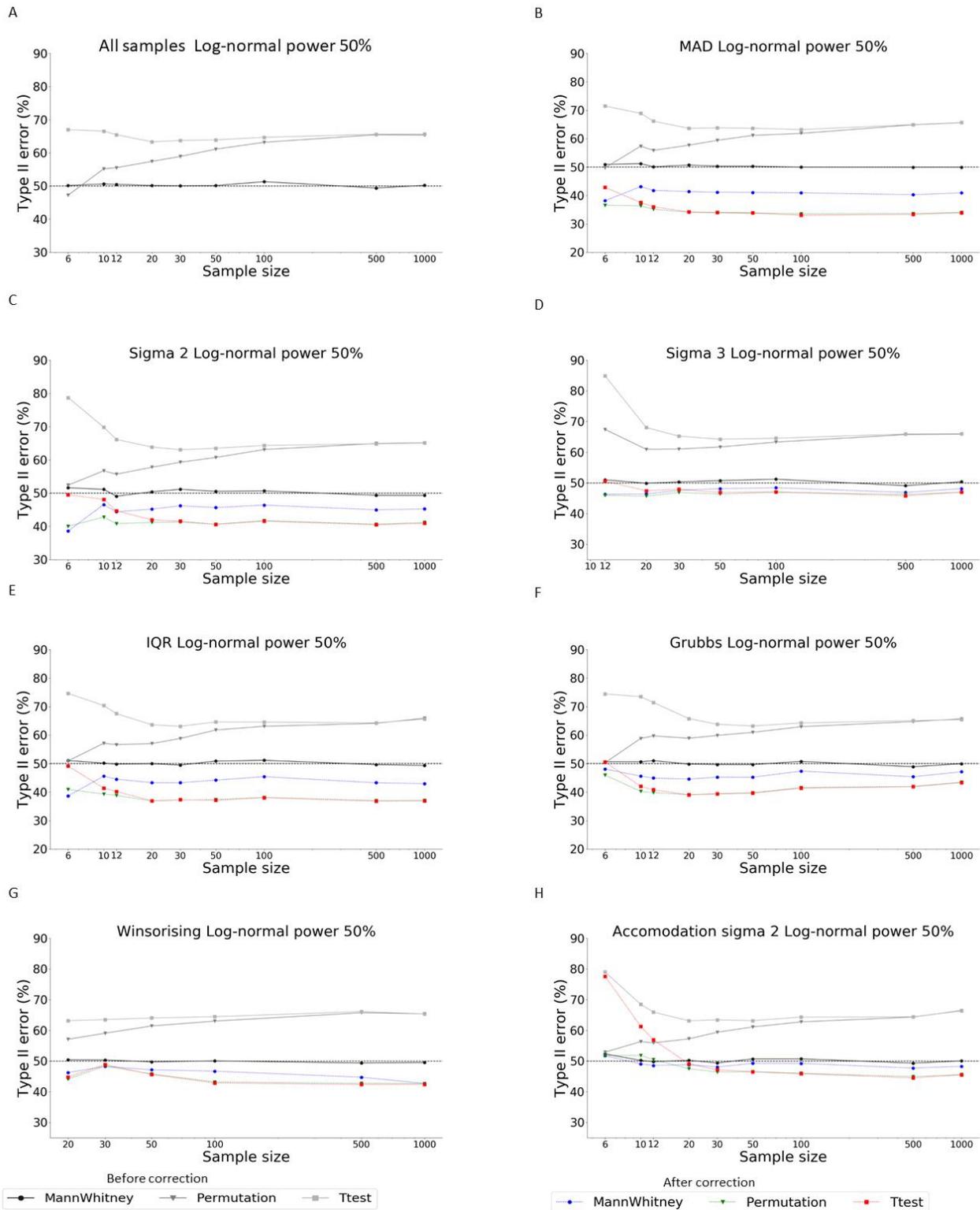

**Fig 12: Impact of correction of "Random sampling outlier" on Type II error when statistical power = 50% (data from log-normal distributions).** In A samples are not selected based on the presence of outliers. All data can be found at https://osf.io/gzra8/

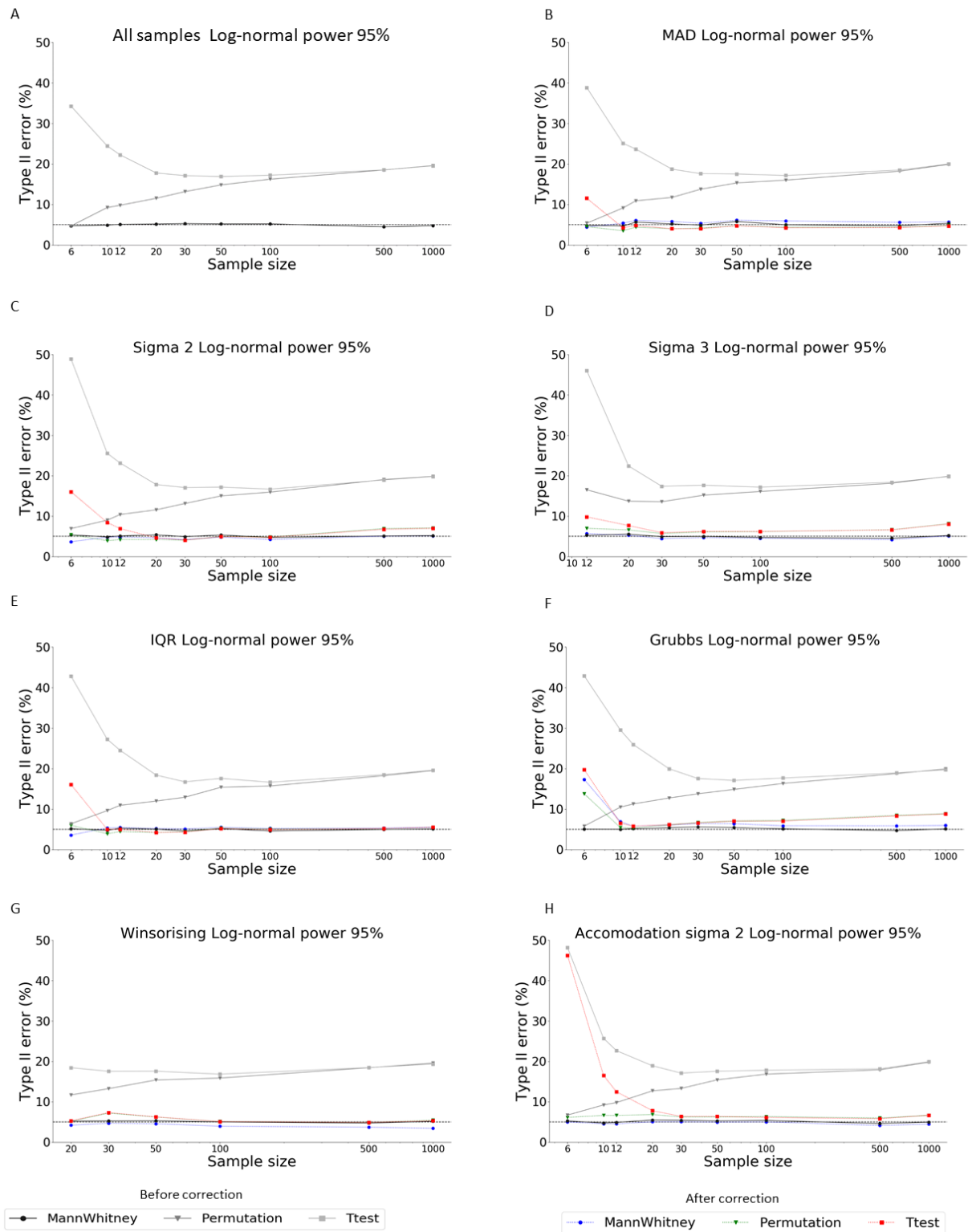

**Fig 13: Impact of correction of "Random sampling outlier" on Type II error when statistical power = 95% (data from log-normal distributions).** In A samples are not selected based on the presence of outliers. All data can be found at https://osf.io/g2xbr/.

*Conclusions from simulation 3*

All together the results from simulation 3 show that:

1- While using non-parametric test with normal distribution do not considerably impact the occurrence of Type II error, parametric and permutation test associated to log normal distribution produce a strong inflation of false negative results.
2- The presence of RSO, detectable with the tested methods, do not modify the probability to observe false negatives results (that remain at the target β risk).
3- Outliers correction methods mainly lead to a reduction of Type II error especially from samples belonging to log normal distribution

4- The reduction of Type II error produced by RSO correction is observed for conditions that increase the risk of Type I error in the absence of real effect (compare figures 7 and 8 with figures 10,11,12 and 13).

## Simulation 4. Impact of RSO on the estimation of real effect

Simulation 4 was performed by the procedure described in Figure 14 (the script is available at https://osf.io/nykbm/). This allows to extract, for the different detection methods, 10000 pairs of samples of sizes **n**, belonging to the separate populations and having at least one RSO in them. The impact of outlier correction on the estimation of population effect was quantified by comparing the difference of the populations means ($\mu_1 - \mu_2$) with the difference of samples means ($\bar{X}_1 - \bar{X}_2$). More precisely, the error on effect estimation (EEE) is calculate as the absolute deviation of sample effect from population effect (EEE=abs(($\mu_1-\mu_2$)- ($\bar{X}_1-\bar{X}_2$)). In these simulations a statistical power of either 50% or 95% was obtained by the modification of the distributions mean ($\mu_2$) from which sample 2 was extracted as a function of the required sample size and distribution types (table 1).

The objective of simulation 4 is to determine whether outlier correction modify the estimation of population effect that can be drown from the samples.

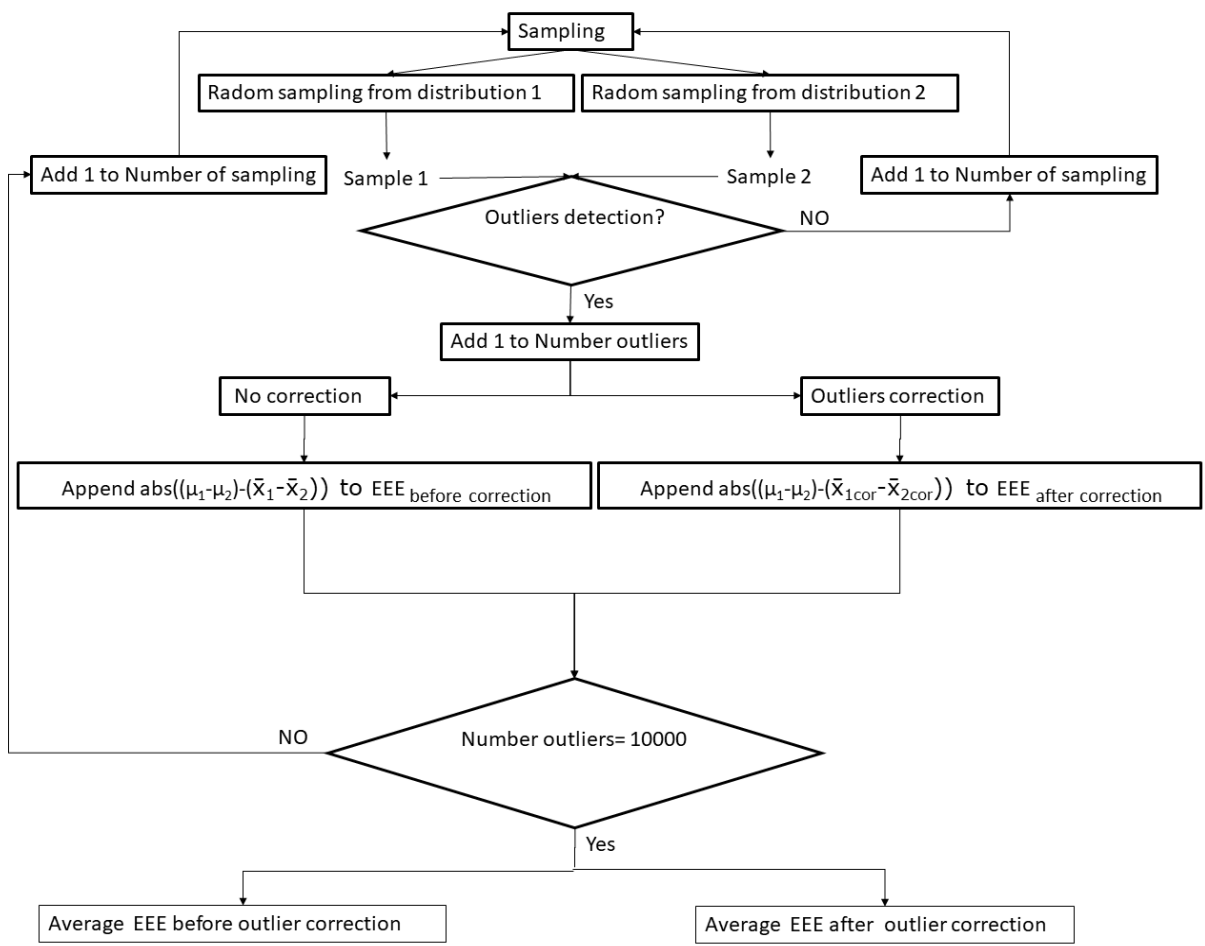

EEE= error in the estimation of the real effect

$\mu$= population mean

$\bar{X}_{cor}$ = sample mean after correction

$\bar{X}$ = sample mean before correction

**Fig 14: Procedure used to simulate the effect of outlier correction on the estimation of population difference** Outliers where separately detected for sample 1 and sample 2. Detection means at least 1 outlier in either sample 1 or sample 2. This procedure was separately applied for the different methods of outliers detection and different types of population distribution (Normal, Log-normal). Script is available at https://osf.io/nykbm/.

*Correction of RSO increase the error committed to evaluate the population effect from the samples comparison (samples from normal population).*

Comparison of figures 15A and 16A show that the error committed in the estimation of real effect from the samples decrease with statistical power. They also show that EEE, expressed as % of modification from real effect, increase with the sample size. It is important to outline that the absolute EEE do indeed decrease with sample size (see data at https://osf.io/v4d9f/), but to lesser extent that the decrease of real effect needed to keep the statistical power at 50% or 95% (see Table 1). This has the consequence to produce the depicted increase of relative modification whit the increase of sample size. Outlier correction produce an increase of the EEE for the majority of the tested methods (Fig 15-16 B-F) except for the "Winsorizing" and "Accommodation Sigma 2", methods for which no modifications are observed (Fig 15-16 G-H). These results are observed for both 50% and 95% statistical power.

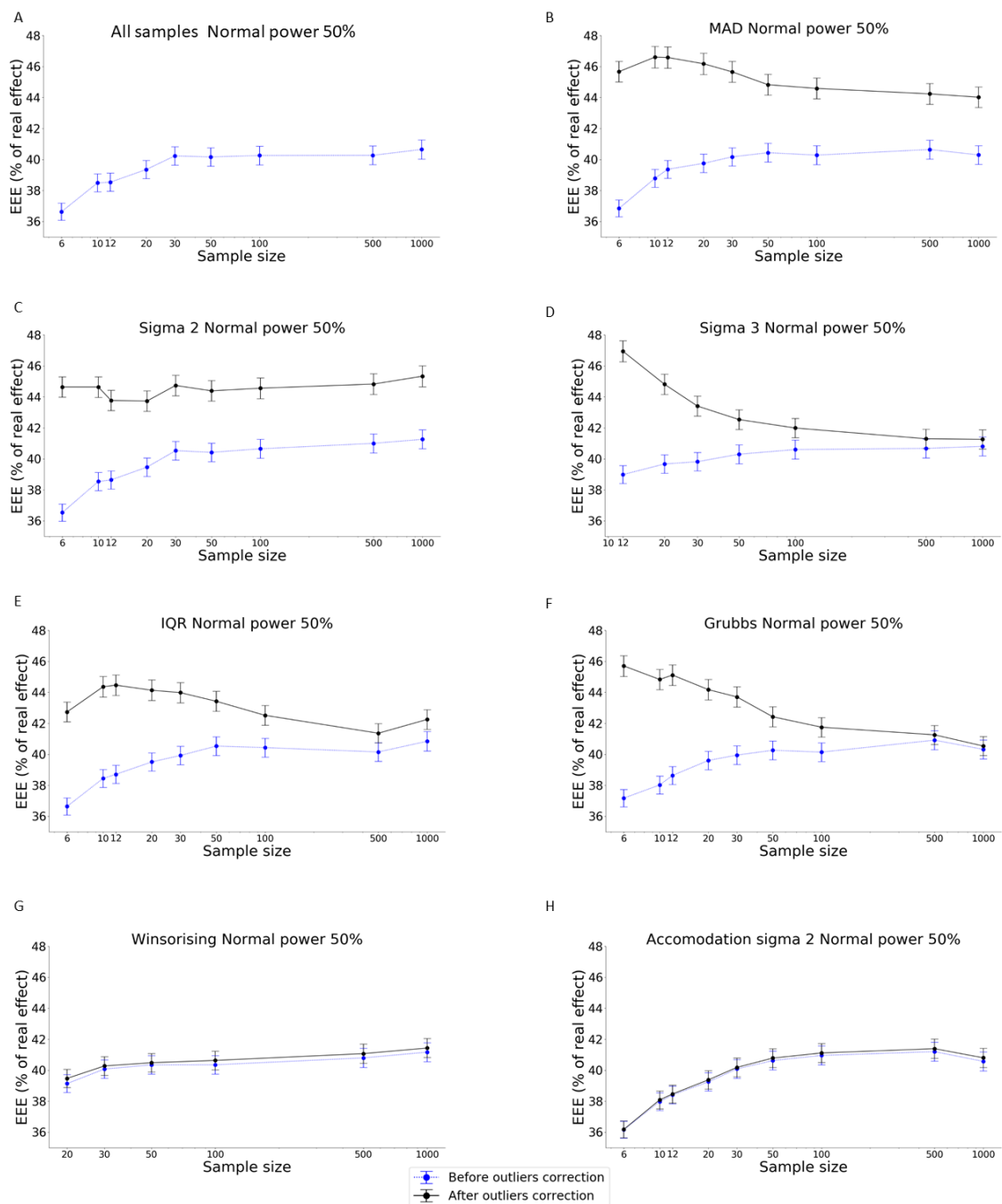

**Fig 15: Impact of correction of "Random sampling outlier" on the estimation of population difference when statistical is power = 50% (data from normal distributions).** EEE= error in the estimation of real effect. In A samples are not selected based on the presence of outliers. Vertical bars represent 95% CI. All data can be found at https://osf.io/v4d9f/.

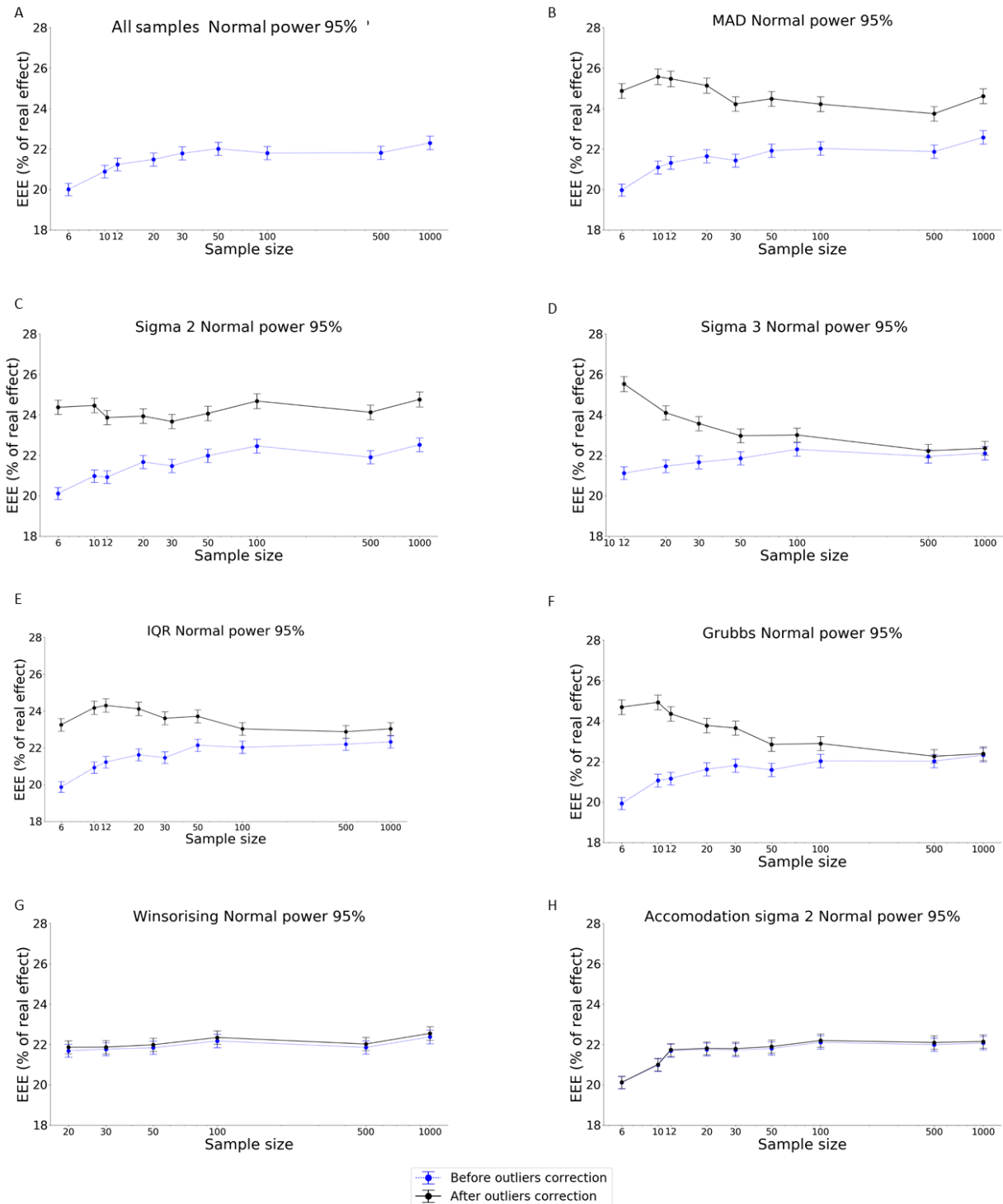

**Fig 16: Impact of correction of "Random sampling outlier" on the estimation of population difference when statistical is power = 95% (data from normal distributions).** EEE= error in the estimation of real effect. In A samples are not selected based on the presence of outliers. Vertical bars represent 95% CI. All data can be found at https://osf.io/xnb7r/

*Correction of RSO mainly decrease the error committed to evaluate the population effect from the samples comparison (samples from normal population).*

When sample belongs to log-normal population the error committed in the estimation of real effect is slightly higher than for sample belonging to normal population. The presence of RSO slightly increase the EEE only when RSO are detectable with Sigma 3 method and sample size is small (compare Fig 17A and 18 A with 17D and 18 D blue lines, sample size=12). Impact of outlier correction change as function of statistical power and correction method. For low power the estimation error on effect quantification is reduced by outlier correction, whatever the method used (Fig 17). However, for comparisons in high statistical power condition the EEE is slightly reduced when "Sigma 3", "Sigma 2", "Accommodation sigma 2" and "Winsorizing" methods are used and increase with the other tested methods (Fig 18).

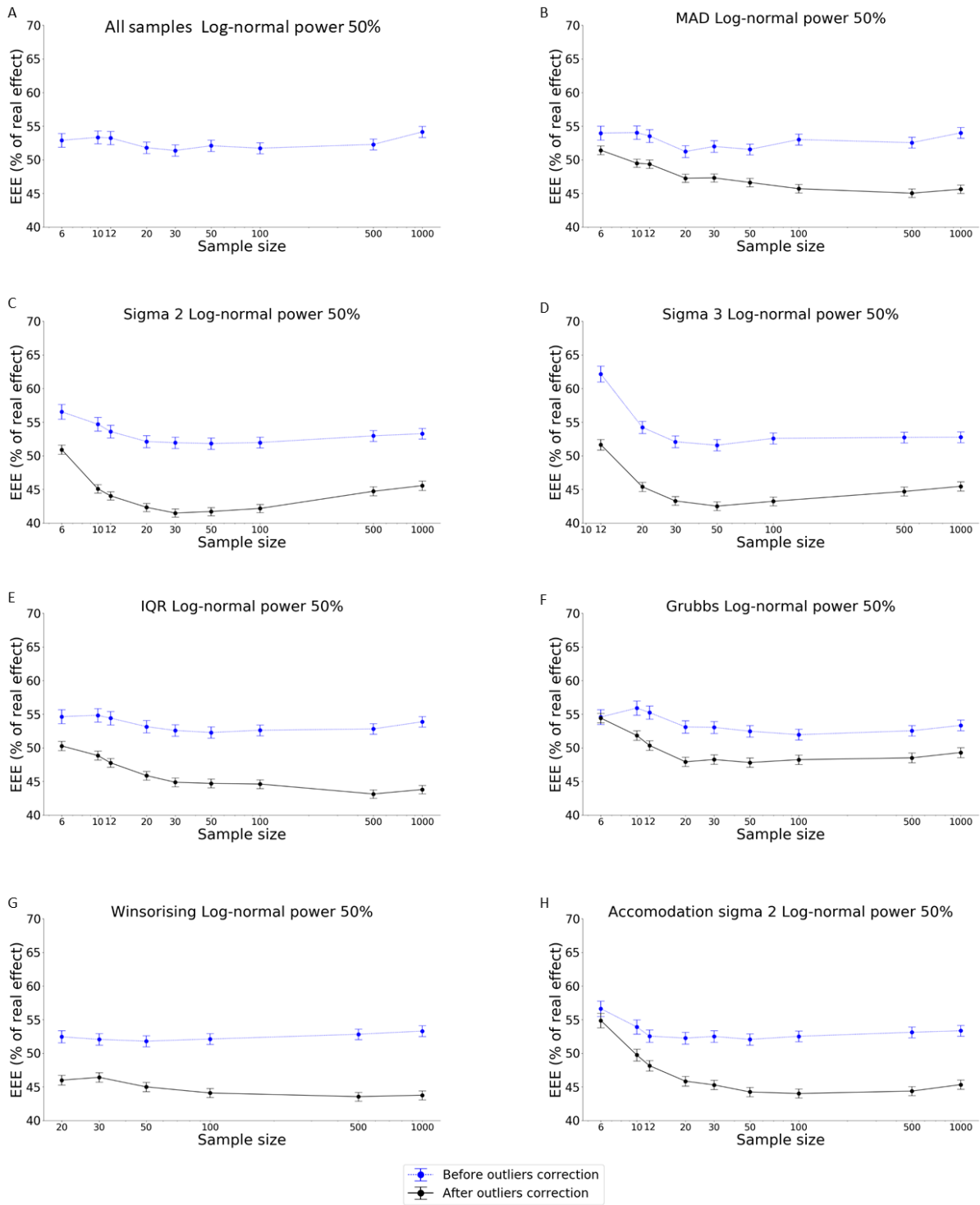

**Fig 17: Impact of correction of "Random sampling outlier" on the estimation of population difference when statistical is power = 50% (data from log-normal distributions).** EEE= error in the estimation of real effect. In A samples are not selected based on the presence of outliers. Vertical bars represent 95% CI. All data can be found at https://osf.io/yrgwk/

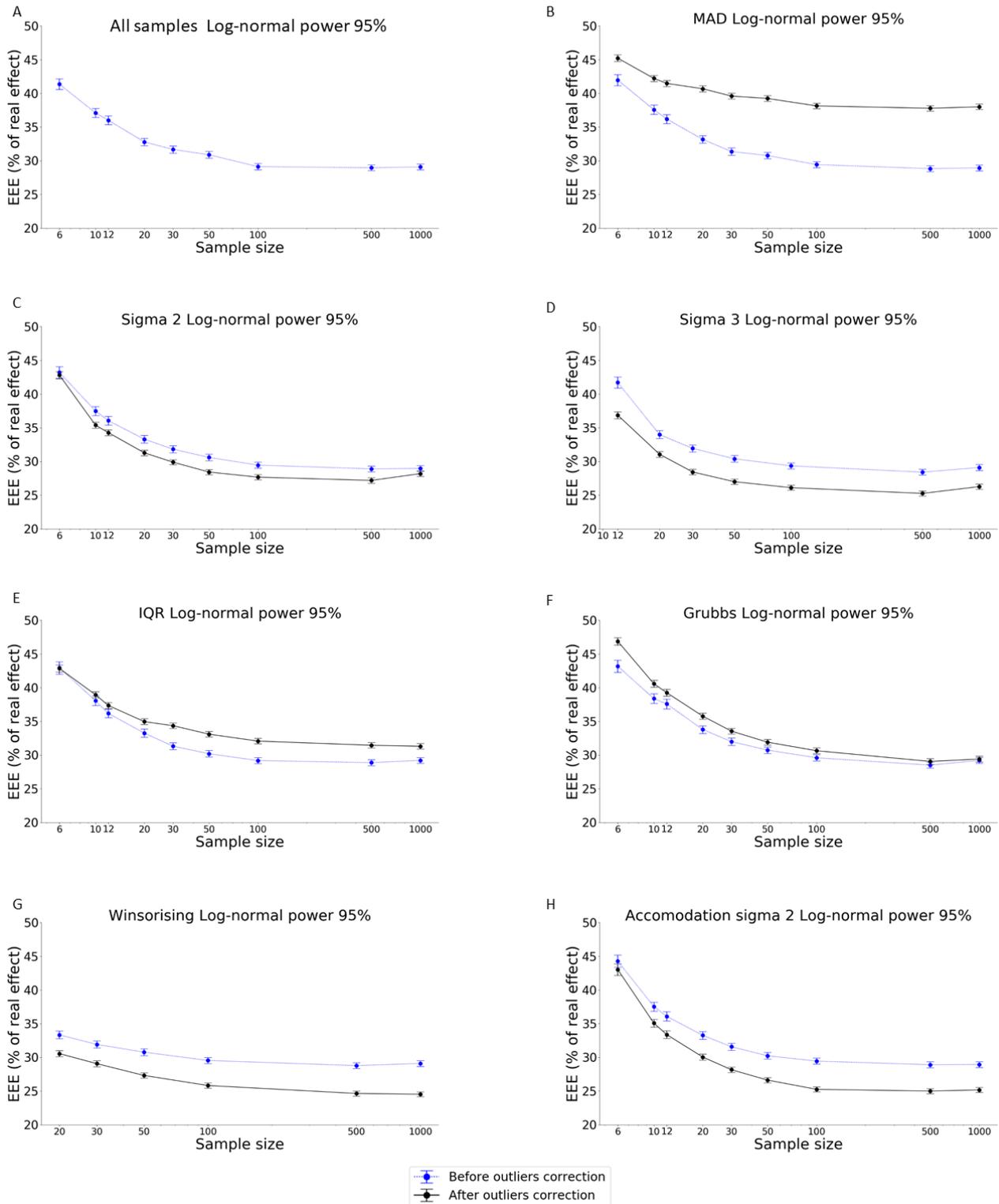

**Fig 18: Impact of correction of "Random sampling outlier" on the estimation of population difference when statistical is power = 95% (data from log-normal distributions).** EEE= error in the estimation of real effect. In A samples are not selected based on the presence of outliers. Vertical bars represent 95% CI. All data can be found at https://osf.io/2w8df/

*Conclusions from simulation 4*

All together the results from simulation 4 show that:

1- When data come from normally distributed population, outliers correction methods applied to RSO worsens the estimation real effect that can be drawn from the samples, with the exception of winsorizing and accommodation sigma 2 methods.

2- When data come from population log-normally distributed, outliers correction methods applied to RSO improve the estimation real effect that can be drawn from the samples, if sampling is carried in low powered condition but worsen the EEE in higher powered condition when MAD, IQR and Grubbs methods are used.

## Simulation 5. Use of outlier correction methods for p-hacking

As shown before (simulation 2) outlier correction in the absence of real outliers increase the probability of type I error for most of the methods tested. Combination of outliers correction methods could therefore potentially be used as strategy for data dredging. To test the impact of this strategy on the occurrence of Type I error we used the procedure descripted in Figure 19 (the script is available at https://osf.io/7s3gv/).

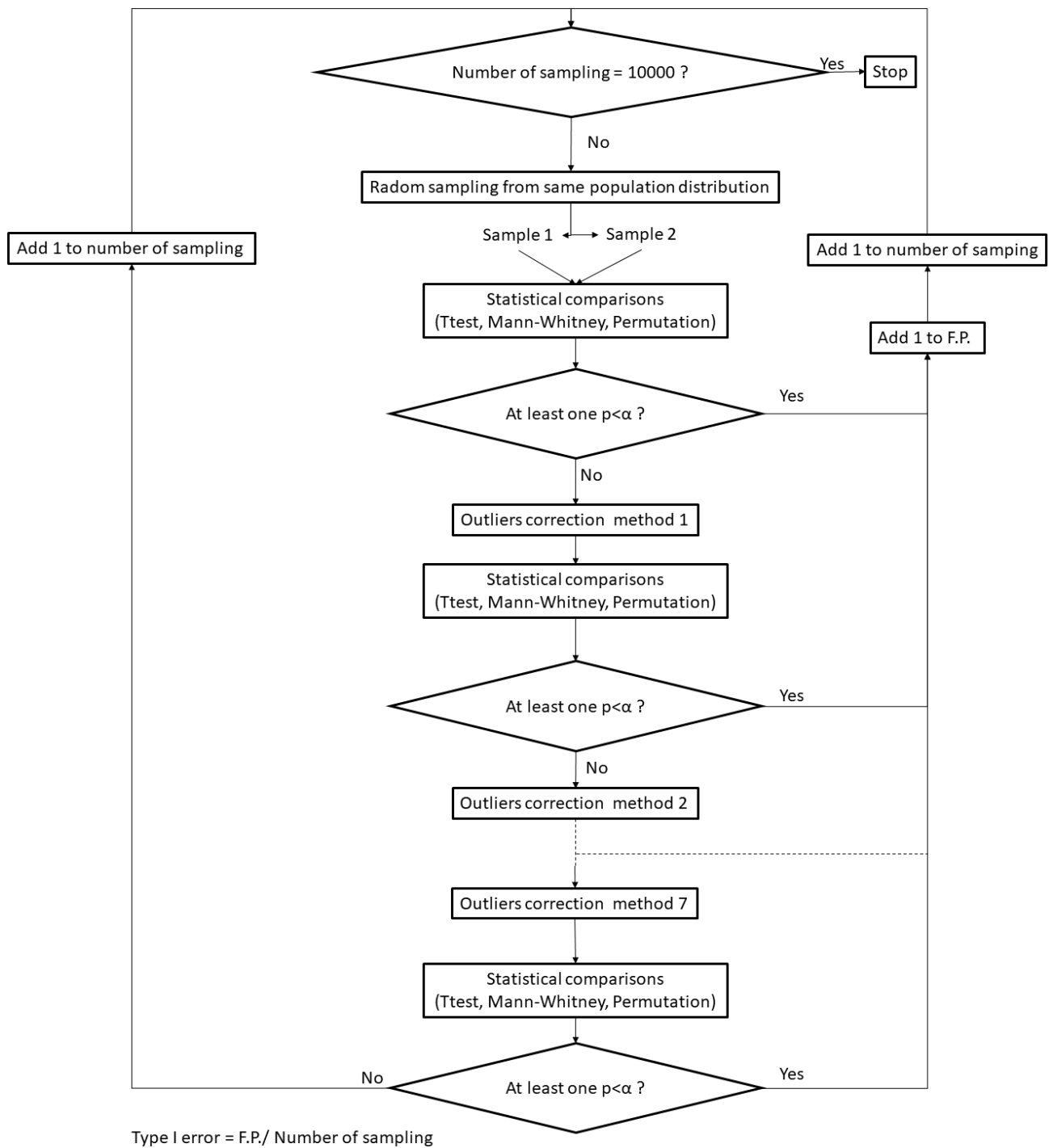

Type I error = F.P./ Number of sampling

**Fig 19: Procedure used to simulate the impact on type I error produced by the utilisation of outliers correction methods in a data dredging strategy.** Seven correction methods are used ( MAD, Sigma 2, Sigma 3, IQR, Grubbs, Winzorizing and Accomodation sigma 2). F.P.= False positives results. "Number of sampling ", "Number of outliers "and "F.P." were set to zero at the beginning of the simulation (script is available at https://osf.io/7s3gv/)

As shown in figure 20 Type I error is strongly inflate when outliers correction is used with a the p-haking strategy. The inflation is greater when samples come from log-normal distribution becoming close to 50% for large sample size.

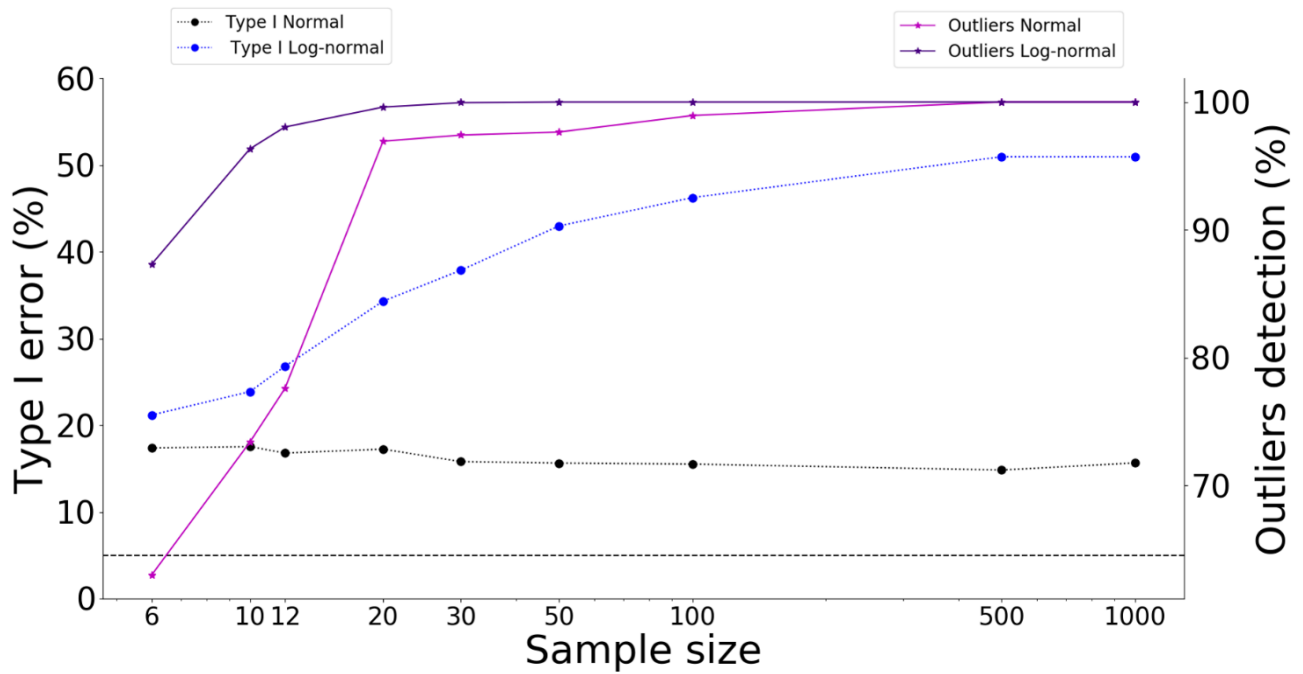

**Fig 20: Inflation of Type I error produced by the utilisation of outlier correction methods through a data dredging strategy.** Samples from normal and log normal distributions. Dashed line represent the σ risk (5%). All data can be found at https://osf.io/xm3nu/

## Conclusions from part 1

The main conclusions are that the presence of extreme values belonging to the studied population, i.e. the RSO, do not or only slightly impact the inferential evaluation of real population from the sample. In particular, for what concern the evaluation of population parameters (mean and STD), Type I and Type II error and real effect estimation. On the other hand, the correction of RSO by most of the methods tested in this study worsen the inferential evaluation of the studied population by increasing the error in the estimation of population parameters (simulation 1) and Type I error (simulation 2). In several condition corrections of RSO can drop the occurrence of Type II error below the expected β risk (simulation 3), however such decrease is often associate with a stronger error in the evaluation of the real effect, in particular for data coming from normal distribution (simulation 4). The impact of the correction change with the method used, with accommodation methods (accommodation sigma 2 and winsorizing) having the lowest negative impact on parameter estimation, Type I error and real effect evaluation when associated to Man Whitney test but not producing reduction of Type II error rate. On the other hand, method having the larger beneficial effect in Type II error (MAD, Sigma 2, IQR) also produce the stronger inflation of Type I error. In general, the negative impact of outlier correction is reduced when statistical comparisons are performed with Mann Whitney test. Finally, our results show that the improper use of outliers correction methods for data dredging can bring to a strong inflation of false positive results (simulation 5).

## Part 2: Contaminating outliers

The evaluation of the effect produced by sample contamination with data which do not belong to the studied population (contaminating outliers), was performed through simulations on which extreme values were injected in the analysed sample (similar to the approach used by (Bakker & Wicherts, 2014; Liao *et al.*, 2016, 2017). In practice "**n**" data of one of the two samples were substituted by "**n**" outliers randomly selected from uniform distribution situated between 4 and 8 (simulation of Type I error) or 4 and 5 (simulation of Type II error) STD of the original distribution. Where **n**= 0,1,2,3,4,5,6,7. When Type II error was evaluated, outliers were injected in the sample coming from the distribution having **µ**=0. In this way increasing false negative rate.

## Simulation 6. Impact of injected outliers and their correction on the estimation of population parameters

This simulation was performed by the procedure depicted in figure 1 with the exception that sample was contaminated with the injected outliers and 200000 sample were analysed regardless the fact that outliers were detected. Only results obtained for sample size equal to 20 are presented. Readers interested to run the simulation with other parameters are invited to use the script available here (https://osf.io/quzm4/).

As expected outlier injection increase the error committed in estimating population mean and STD inferred from sample mean and STD (Fig 21 black triangles). All the outlier correction methods tested reduce the error generated by outlier injection when sample came from normal distribution (Fig 21 A-B). A stronger efficacy was observed, in the order, for MAD, IQR, Sigma 2 and Grubbs methods. Similar effect are observed when sample come from log-normal distribution with the exception that when only one outlier is present MAD, Grubbs and IQR correction increase the error in STD estimation (Fig 21D).

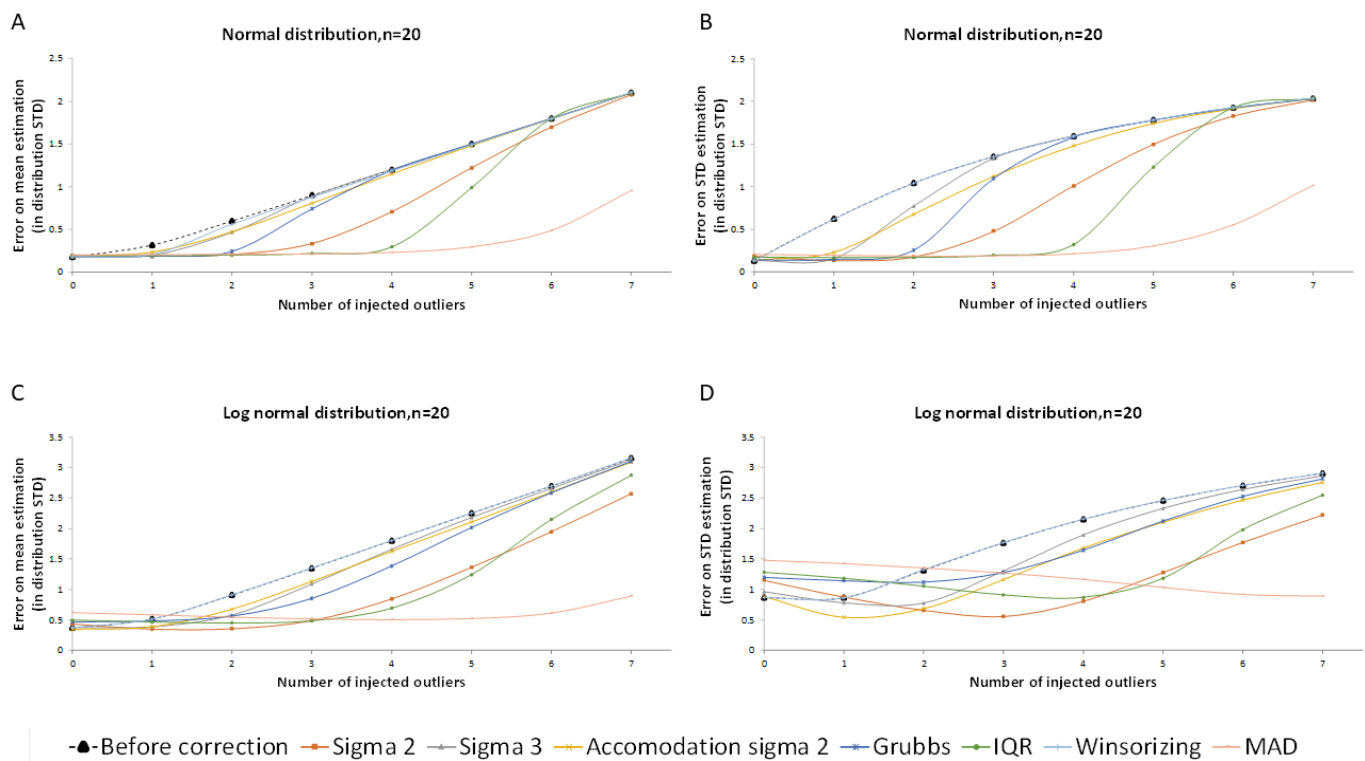

**Fig 21: Impact of outlier injection and their correction on μ and σ estimation error.** All data can be found at https://osf.io/v6kjd/

## Simulation 7. Impact of injected outliers and their correction on Type I error

This simulation was performed by the procedure depicted in figure 6 with the exception that sample 1 was contaminated with the injected outliers and 200000 paired of sample were analysed regardless the fact that outliers were detected. Only results obtained for sample size equal to 20 are presented. Readers interested to run the simulation with other parameters are invited to use the script available here https://osf.io/quzm4/).

As expected outlier injection in one of the two samples increase the occurrence of Type I error for samples coming both from normal and log-normal distribution (Fig 22 A-D). When sample come from normal distribution outliers correction counteract the inflation of Type I error produced by outlier injection, but only when MAD, IQR and Sigma 2 methods are used (Fig 22A-B). MAD method indubitably showing the higher efficacy. It is important to note that the beneficial effect of outlier correction is present only when the number of outliers in each sample is larger than two (more than 20% of the data are contaminating outliers). On the other hand, when samples come from Log-normal distribution outlier correction further increase the inflation of Type I due to outlier injection. This counterintuitive result, illustrated by the examples in Figure 22 E and F, appears to be due to the fact that contaminating outliers tend to be considered normal data with respect to the sample while at the same time RSO are detected in the sample not having injected outliers.

It's important to note that the inflation of Type I error produced by outlier injection is strongly reduced when samples are compared with non-parametric test with respect to the use of parametric test (compare black triangles in A Vs B and C Vs D).

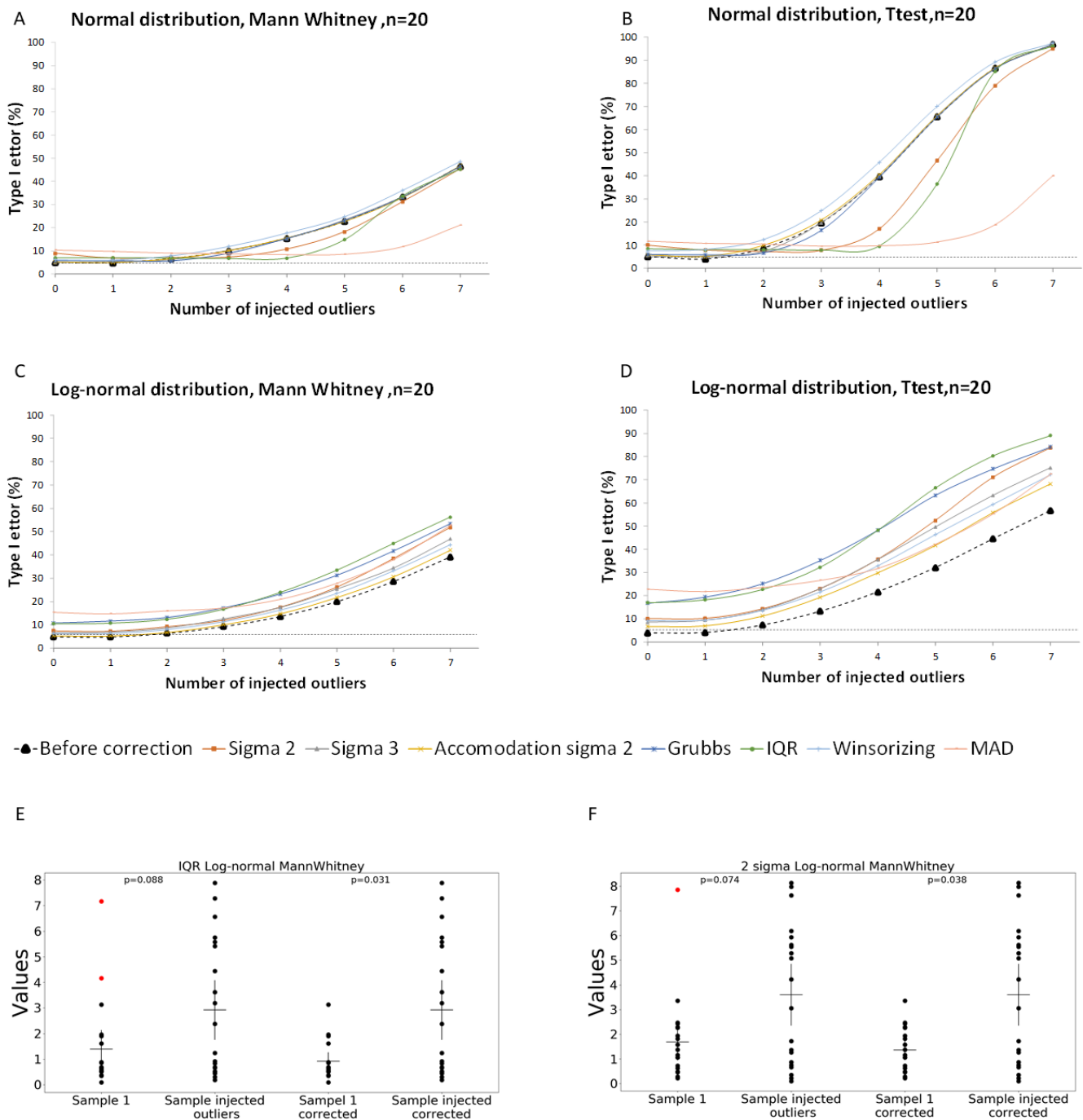

**Fig 22: Impact of outlier injection and their correction on Type I error.** A-D) Type I error as a function of the number of injected outliers, population distribution and statistical test. E-F example of outliers correction that producing false positive results. Red dots represent data detected as outliers. 7 outliers injected, n=20. Dashed line All data can be found at https://osf.io/2bcdy/. Script for generation of figure E, F can be found at https://osf.io/g39j2/.

# Simulation 8. Impact of injected outliers and their correction on Type II error

This simulation was performed by the procedure depicted in figure 9 with the exception that sample 1 was contaminated with the injected outliers and 200000 paired of sample were analysed regardless the fact that outliers were detected. The results presented are obtained in a condition in which sample size equal to 20 and statistical power in the absence of outliers is 95%. Readers interested to run the simulation with other parameters are invited to use the script available here https://osf.io/8ntwy/).

As expected outlier injection in one of the two samples increase the occurrence of Type II error for samples coming both from normal and log-normal distribution (Fig 23 A-D). Again it should be noticed that this inflation is lower when samples are compared with Mann Whitney test (black triangles). When samples came from normal distribution outlier correction counteract the inflation of Type II error associated to outliers injection. Particularly when MAD, IQR and Sigma 2 methods are used. The relative effect of outlier correction is stronger when samples are compared with the parametric test. But this is due to the fact that the Type II inflation in the absence of correction is bigger when Ttest is used. Indeed, the absolute value of Type II error after outliers correction is lower for Mann Whitney than Ttest condition, (use horizontal lines in Fig 23 A and B to compare the Type II error after correction when injected outliers are 3 or 4). When samples came from a Log-normal distribution, outlier correction increase Type II error inflation when statistical comparison is performed with Mann Whitney (except for winsorizing method) but is reduced when performed with Ttest. However, the absolute value of type II error after correction is similar for both type of tests with the exception of the correction with Accommodation sigma 2 and Winsorizing methods for which Type II error, when evaluated with the Ttest, is lower (use horizontal lines in Fig 23 C and D to compare methods when injected outliers are 2 or 5).

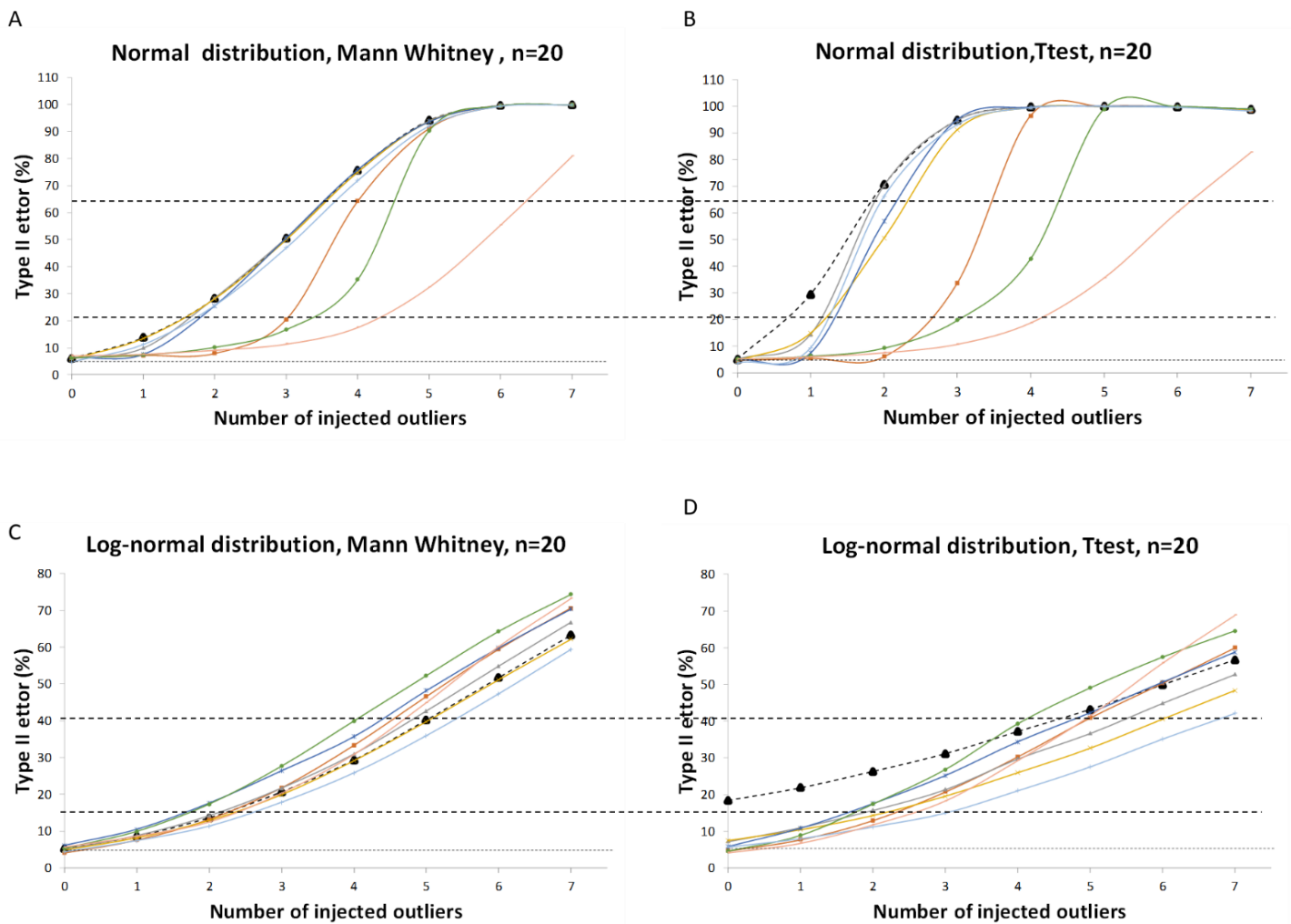

**Fig 23: Impact of outlier injection and their correction on Type II error .** A-D) Type I error as a function of the number of injected outliers population distribution and statistical test. The statistical power in the absence of outlier is 95%All data can be found at https://osf.io/26qb5/

The results of simulation 7 and 8 show that, when sample size =20, the correction methods that produce the larger benefit in the presence of contaminating outliers are also those that produced the higher inflation of Type I error in their absence (MAD, IQR and Sigma 2 ; Fig 7 and 8). In order to look for the existence of experimental conditions for which Type I error is maintained at ~ σ risk in absence of outliers and inferential statistic still benefit for correction of contaminating data, we tested Sigma 3 outlier method for sample size ≥500 (a condition for which no inflation of Type I error is observed in the absence of outliers, fig 7D and 8D). As show in figure 24 when data came from normal distribution this method efficiently counteract the inflation of Type I and Type II error produced by outliers injection while a small increase of the two errors is produced when samples came from log-normal distributions.

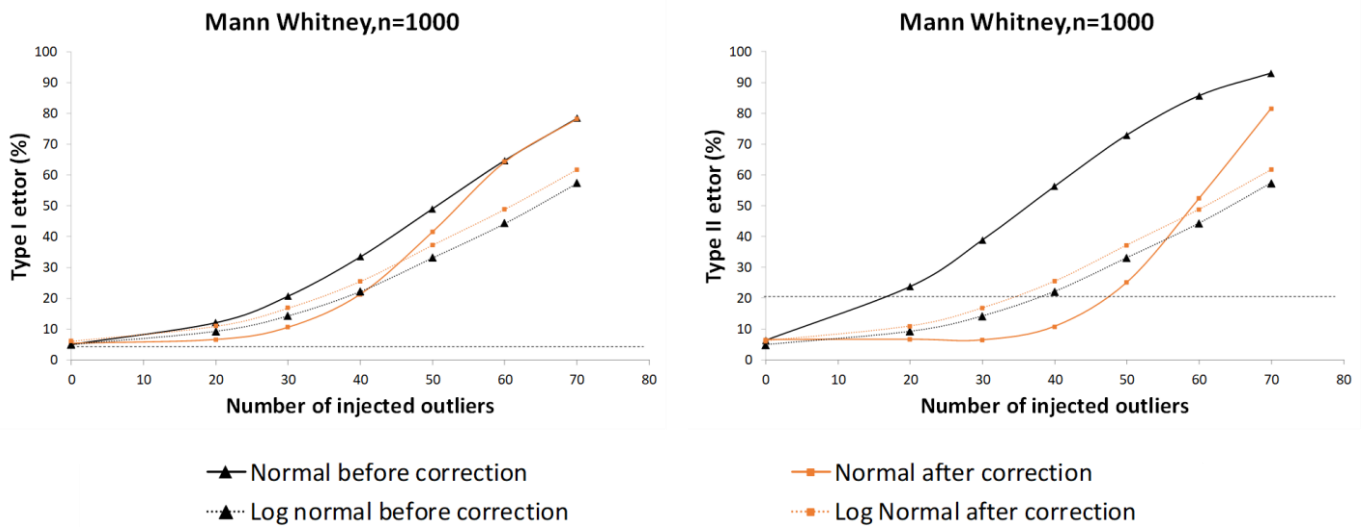

**Fig 24: Impact of Sigma 3 correction on Type I and Type II error inflation produced by outliers injection.**

Up to now the simulations using MAD correction method have been performed using a threshold of 2.24. However recent reports suggest that the more restrictive threshold of 3 should be used instead (Leys *et al.*, 2013, 2019). We therefore compared the impact of outlier correction on Type I error for the two thresholds. As shown in figure 25A and 25B the inflation of Type I error in the absence of contaminating outliers is reduced by the utilization of the more restrictive threshold. However, it remains largely above the σ risk, especially for samples coming from log-normal distribution and lower sample sizes when sample came from normal distribution. On the other hand, increasing the threshold reduced the efficacy of MAD method to counteract the inflation of Type I error produced by contaminating outliers (Fig 25 C and D).

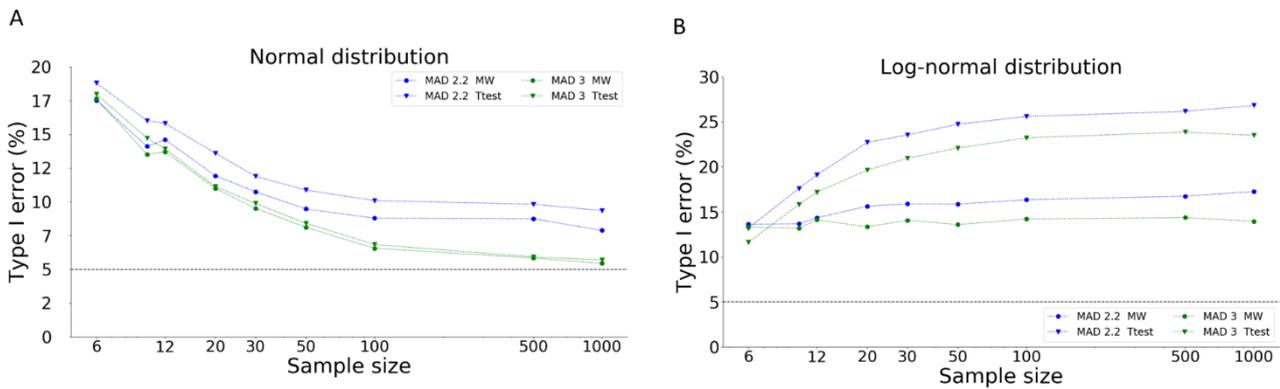

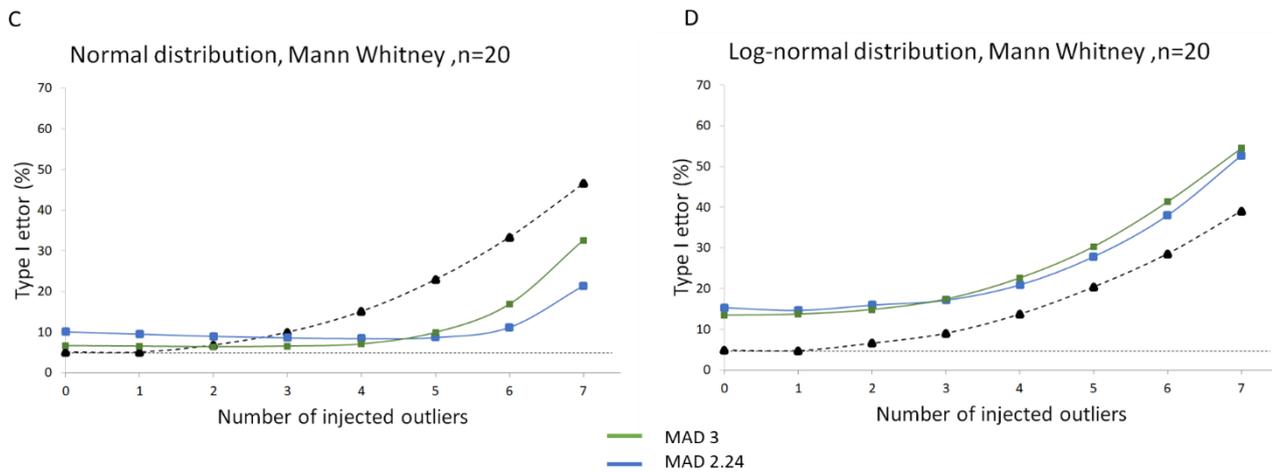

**Fig 25: Comparison of outlier correction on type I error when MAD method is applied with different threshold.** A and B) when sample do not contain contaminating outliers. C and D) in the presence of contaminating outliers. MW= Mann Whitney test

## Conclusions from part 2

The main conclusions are that the presence of extreme values not belonging to the studied population, i.e. the real outliers, negatively impact the inferential evaluation from the sample. All the correction methods tested counteract the increase of the error committed in the evaluation of population mean and STD that is produced by outliers injection; with MAD, IQR and Sigma 2 methods having the highest efficacy. This irrespectively of the fact that sample came from normal or log-normal distribution (simulation 6). Both before and after outlier correction the inflation of Type I end Type II error produced by outliers injection is lower when samples are compared with non-parametric (Mann Whitney) than parametric (Ttest) test (except the particular case of accommodation methods for Type II error in log-normal distributions, Fig 23D). When sample size=20 only MAD, IQR and Sigma 2 methods are able to counteract the inflation of Type I and Type II errors produced by outlier injection, however their efficacy is observed only when sample came from normally distributed population, being often responsible of a further increasing of false positive or false negatives results when data are sampled from log-normal distribution (simulation 7 and 8). The methods that show the larger benefit in the presence of contaminating outliers (MAD, IQR and Sigma 2) are also those that produced the higher inflation of Type I error in the absence of contaminating outliers (Fig 9 and 10).

## Discussion

As shown in the present report, the inefficacy of common methods of outliers correction stems from the fact that none of these is capable to distinguish outliers data belonging to the studied population, i.e. the RSO, from contaminating data. Since RSO occurs with high probability and since their removal worsen the inferential evaluation of the studied population, the systematic utilisation of outliers correction risk to produce more harmful than beneficial effect on statistical inference. This is particularly true for what concern Type I error that is considerably increased by the removal of RSO. This result is in agreement end extend what previously report by Bakker and Wicherts 2014 that found an increase of Type I error when methods based on mean plus or minus sigma were applied to samples coming from normal distribution. In their report however the inflation of false positive results was slightly lower than what reported here and increased whit the sample size instead of decreasing. This discrepancy is likely due to the fact that in their simulation were also included samples not contains RSO, therefore the inflation Type I error depend both on the fallacy of the correction method as well as on the its probability to detect RSO. We believe that the selective selection of samples presenting RSO, allow a better comparisons of the risk-benefit ratio associated to the different correction methods. Whit this approach we showed that those methods that, in the absence of contaminating outliers, present the lower risk of type I error inflation and the lower risk of inferential evaluation worsening (ex. the Accommodation Sigma 2) also have the lowest efficacy in counteracting Type I and Type II error inflation produced by contaminating outliers. On the contrary those method having the highest efficacy against contaminating outliers (ex. MAD) produce stronger inflation of Type I error in their absence. As suggested by simulation 5, the inflation of false positive results could be further accentuated by the inexistence of specific guidelines on the use of outlier correction methods, that leaves the researcher free to choose or not the method that best fits its own data, or its expectancy on data interpretation (Simmons *et al.*, 2011; Holman *et al.*, 2015). Since their diffuse utilization (Leys *et al.*, 2013; Bakker & Wicherts, 2014) the systematic utilisation of outlier correction procedures investigated here are likely to contribute to the reproducibility crises associated to methodological limits and experimental bias (Chambers, 2017; Munafò *et al.*, 2017),the problem of Type I inflation linked to outlier correction should therefore be explicitly addressed by journal's editorial boards. How to deal with contaminating outlier? As proposed by Leys and coo workers the decision of removing a putative outlier from a data set should not be taken exclusively based on the outcome of a mathematical test (at least not one of those investigated in the present study), "Researchers are recommended to follow a two-step procedure to deal with outliers. First, they should aim to detect the possible candidates by using appropriate quantitative (mathematical) tools. …. Second, they should manage outliers and decide whether to keep, remove, or recode these values, based on qualitative (non-mathematical) information" (Leys *et al.*, 2019). Although in some specific condition mathematical methods appears to have a high reliability to mainly detect contaminating outliers (ex Sigma 3 method for sample size ≥500, see figures 7D, 8D and 24). The simulation approach used in the present reports could be a useful tool to look for conditions not investigated here or for seeking new mathematical methods more resistant to the inflation Type I error produced by removal of RSO. For the moment the most reliable, non-subjective, way to reduce the impact of contaminating outliers appears to be the utilisation of non-parametric test, at least when data come from normal and log-normal distribution. Indeed, as shown in figure 7A, the use of Mann Whitney in the absence of contaminating data keep Type I error at **σ** risk when is used with normally and log normal distributed data, while Type II error is only slightly increase when samples came from normal distribution (Fig 10 A and 11 A). On the other hand, the inflation of Type I and Type II error produced by contaminating outliers is strongly reduced by the use of Mann Whitney test compared to Ttest for samples coming both from normal or log-normal distribution (Fig 22 and 23).

## Acknowledgments:

This work was supported by the CNRS, Inserm, and Lyon 1 University. We wish to thank Christophe Leys for its pertinent comment on the manuscript as well as Nicolas Fourcaud-Trocméand and Samuel Garcia for helping produce the scripts for data simulation.

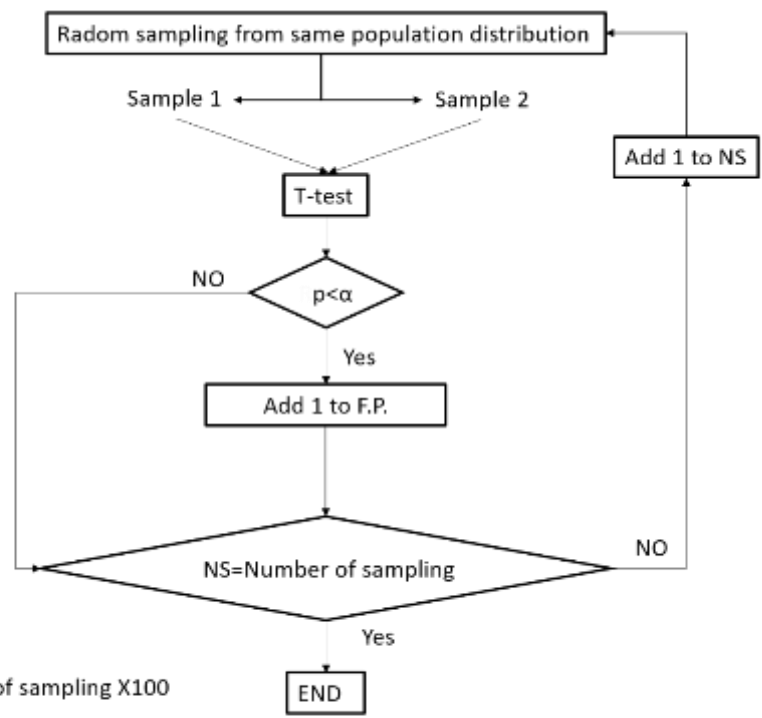
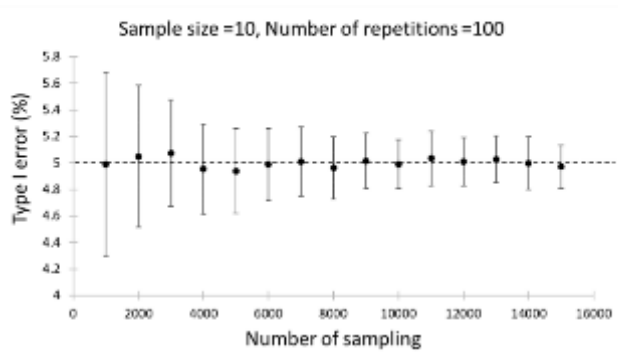
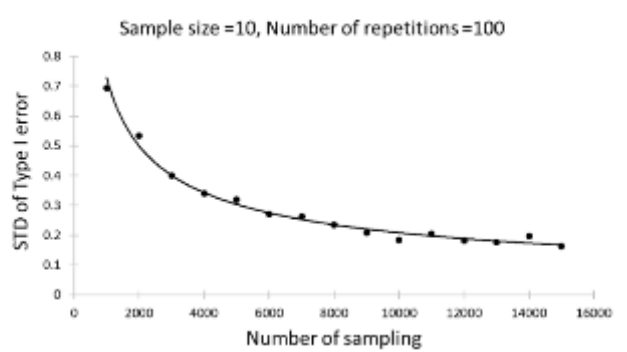

**Supplementary figure 1: Procedure used to empirically define the number of sampling for the simulations. A)** flow chart of the procedure used to estimate Type I error as a function of the number of sampling. **B)** The variability of the estimation of

Type I error as a function of the number of sampling after 100 repetitions of the procedure in A. **C)** Evolution of STD depicted in B. When the number of sampling is 10000, ~95% of the simulations are expected to give a Type I error in the range [4.6, 5.4], i.e. close to the attended **σ** risk.

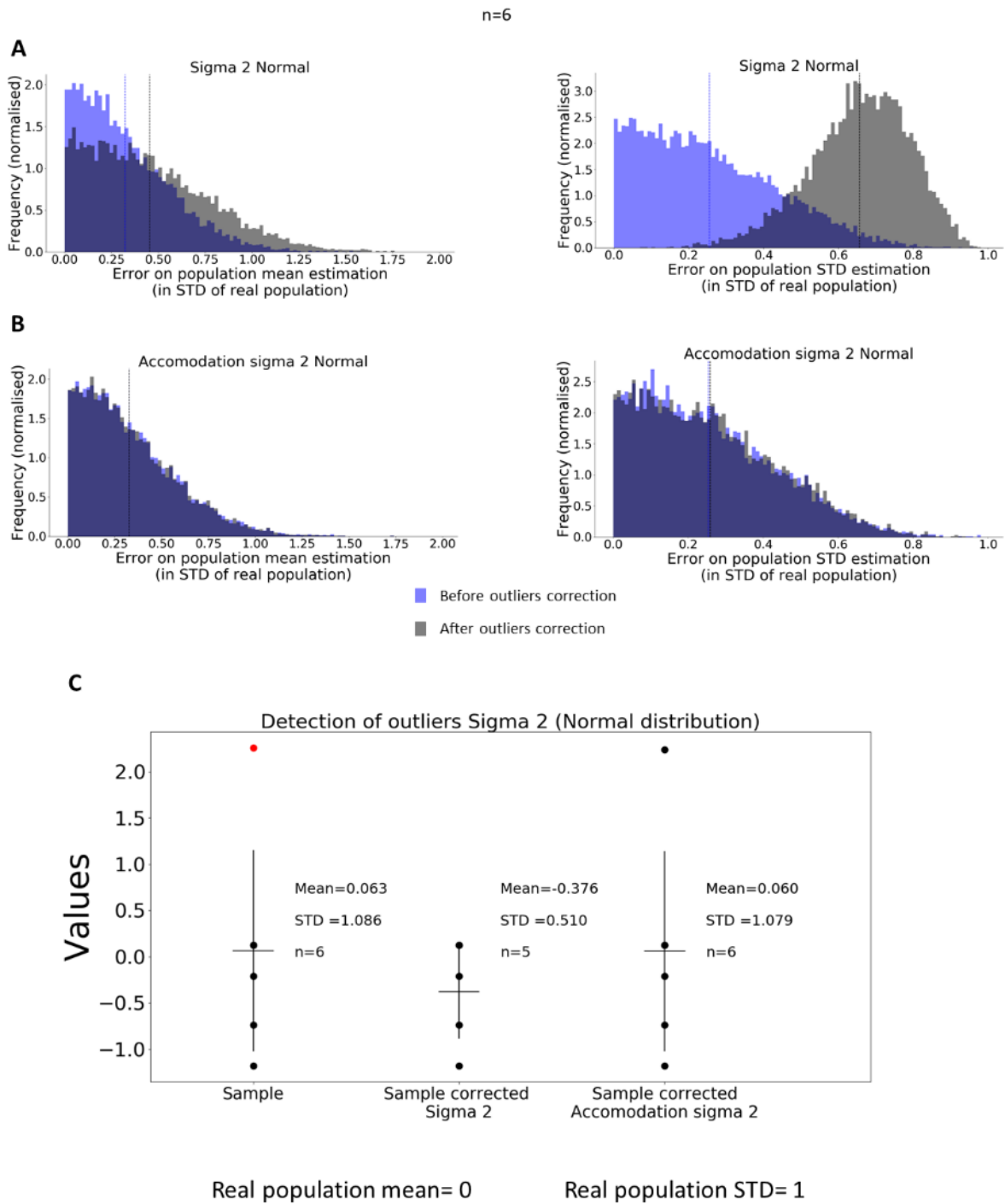

**Supplementary figure 2: A-B** Distribution of **μ** and **σ** estimation errors, before and after outliers correction. **A** When "Sigma 2" method is used. **B** When "Accommodation sigma 2" method is used. **C.** Example of sample correction that brings to an increase of **μ** and **σ** estimation error when "Sigma 2" correction method is used but no relevant change on the estimation of these two parameters when "Accommodation sigma 2" method is used. Sample from normal distribution. Sample size=6. Vertical dot lines represent the means of error distributions. Scripts for simulation A-B and C are respectively available here https://osf.io/dfeas/ and here https://osf.io/t9mvc/.

n=100

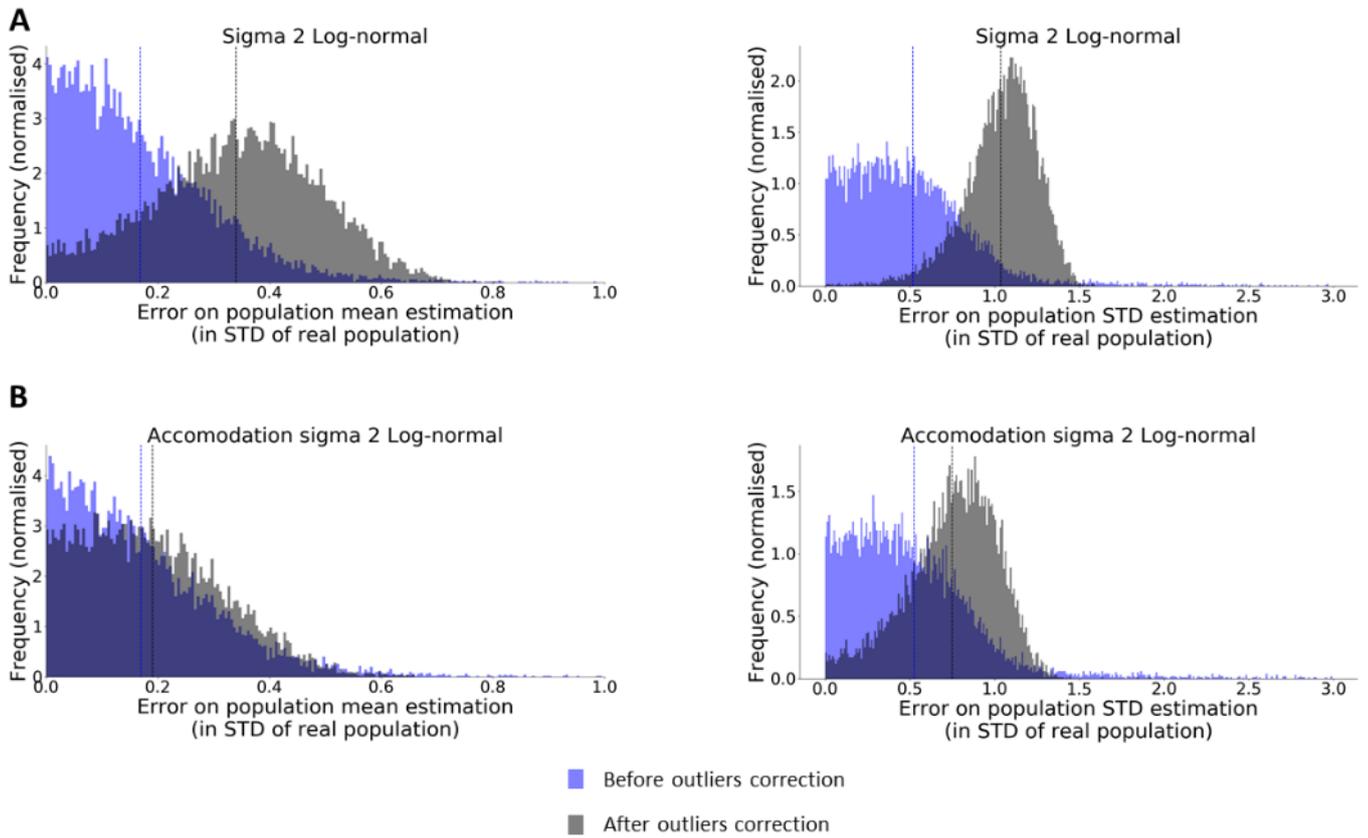

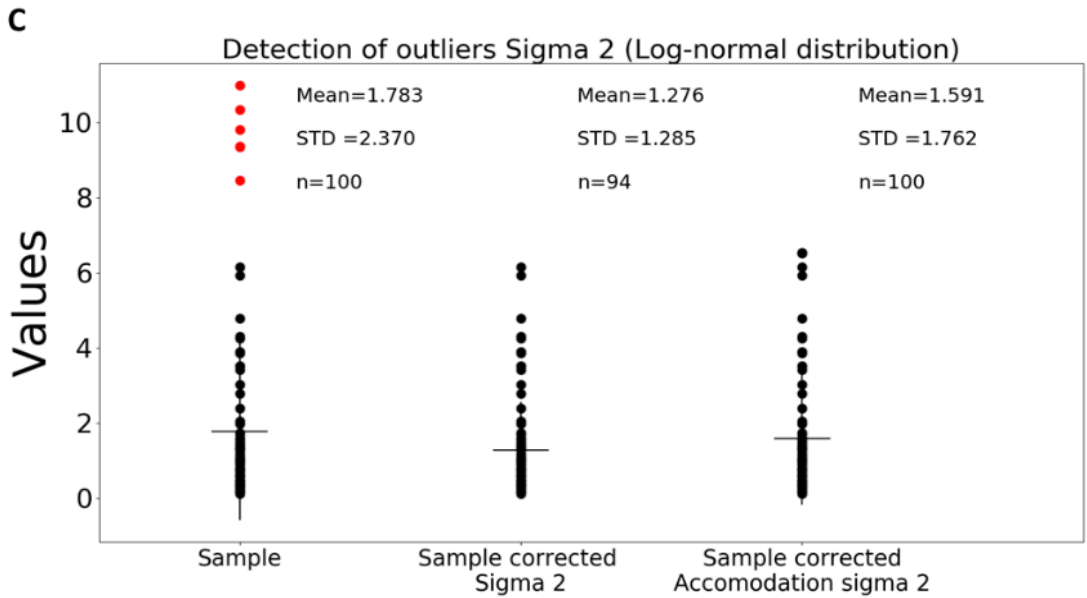

Real population mean= 1.65       Real population STD= 2.16

**Supplementary figure 3:** **A-B** Distribution of μ and σ estimation errors, before and after outliers correction. **A** When "Sigma 2" method is used. **B** When "Accommodation sigma 2" method is used. **C.** Example of sample correction that brings to an increase of μ and σ estimation error when "Sigma 2" correction method is used but no relevant change on the estimation of these parameters when "Accommodation sigma 2" method is used. Sample from log-normal distribution. Sample size=100. Scripts for simulation A-B and C are respectively available here https://osf.io/dfeas/ and here https://osf.io/t9mvc/.

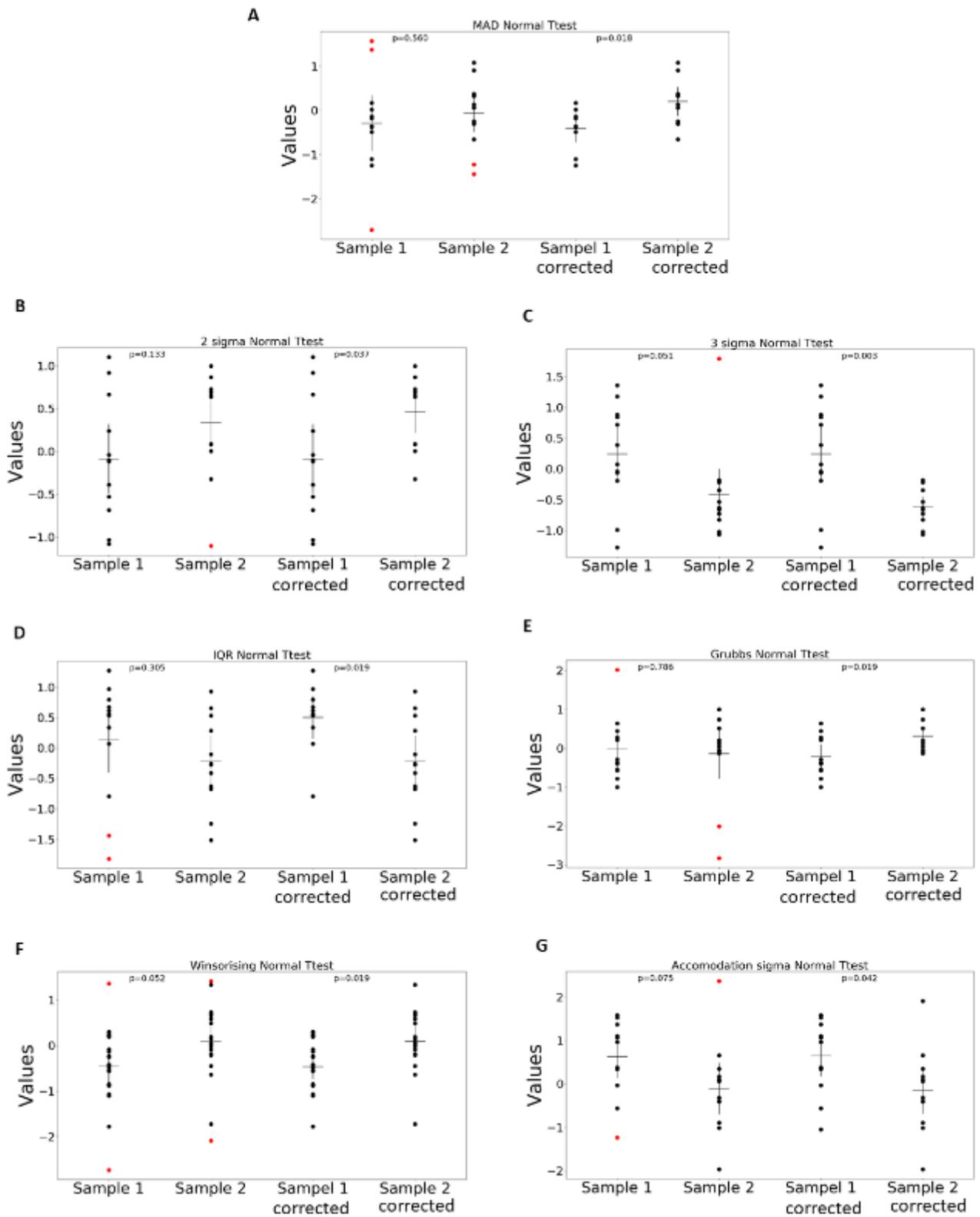

**Supplementary figure 4: Examples of Type I error produced by outlier correction.** Outliers are evidenced by red dots. Note in B and C that values can be detected as outliers, with respect to the sample, even though they are not outliers with respect to the real population (μ=0, sigma=1). The script generating the examples is available at https://osf.io/c529b/.

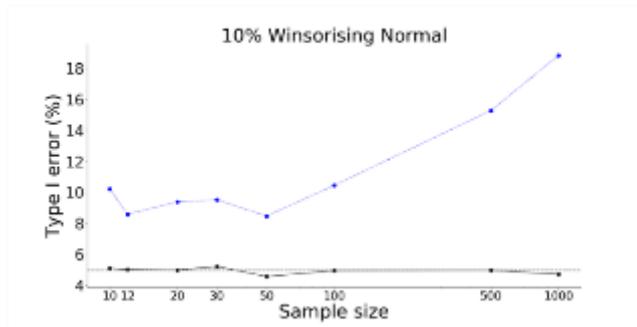
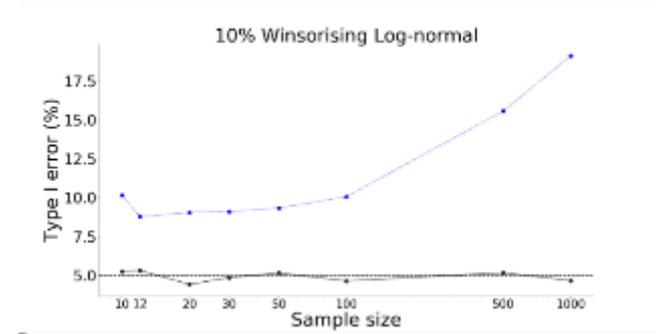
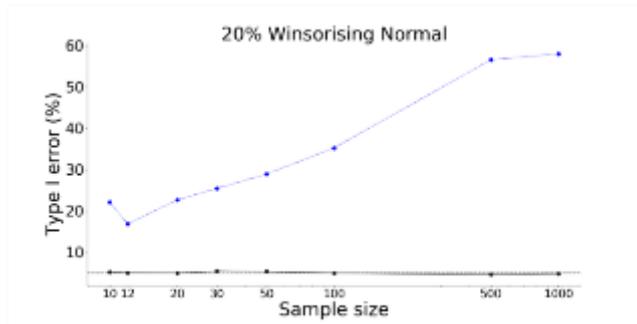
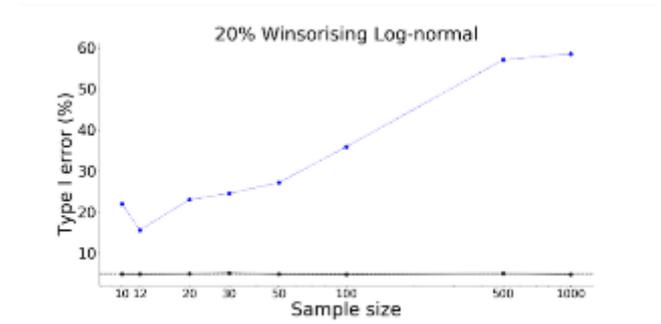

**Supplementary figure 5: Effect of winsorizing at 10% and 20 % on the occurrence of Type I error.** Comparisons were made with Mann Whitney test.